\documentclass[useAMS,usenatbib]{mn2e}

\usepackage{natbib}
\usepackage{graphicx}
\usepackage{color}

\title[The size distribution of Jupiter Family comet nuclei]{The size distribution of Jupiter Family comet nuclei}
\author[C. Snodgrass et al. ]{C. Snodgrass$^{1,2}$\thanks{E-mail:
snodgrass@mps.mpg.de}, A. Fitzsimmons$^{3}$, S. C. Lowry$^{4}$ and P. Weissman$^{5}$\\
$^{1}$Max Planck Institute for Solar System Research, Max-Planck-Str. 2, 37191 Katlenburg-Lindau, Germany\\
$^{2}$European Southern Observatory, Alonso de Cordova 3107, Vitacura, Santiago, Chile\\
$^{3}$Astrophysics Research Centre, Queen's University, Belfast, UK\\
$^{4}$Centre for Astrophysics and Planetary Science, The University of Kent, Canterbury, UK\\
$^{5}$Jet Propulsion Laboratory, Pasadena, California, USA}
\begin{document}

\date{Accepted ; in original form }

\pagerange{\pageref{firstpage}--\pageref{lastpage}} \pubyear{2011}

\maketitle

\label{firstpage}

\begin{abstract}
We present an updated cumulative size distribution (CSD) for Jupiter Family comet (JFC) nuclei, including a rigourous assessment of the uncertainty on the slope of the CSD. The CSD is expressed as a power law, $N(>r_{\rm N}) \propto r_{\rm N}^{-q}$, where $r_{\rm N}$ is the radius of the nuclei and $q$ is the slope. We include a large number of optical observations published by ourselves and others since the comprehensive review in the {\it Comets II} book \citep{Lamy-chapter}, and make use of an improved fitting method. We assess the uncertainty on the CSD due to all of the unknowns and uncertainties involved (photometric uncertainty, assumed phase function, albedo and shape of the nucleus) by means of Monte Carlo simulations. In order to do this we also briefly review the current measurements of these parameters for JFCs. Our final CSD has a slope $q=1.92\pm 0.20$ for nuclei with radius $r_{\rm N} \ge 1.25$ km.

\end{abstract}


\section{Introduction}

The size distribution of any population of solar system small bodies is of critical importance in constraining their formation and subsequent collisional evolution. \citet{Dohnanyi69} showed that a collisionally relaxed population of self-similar bodies, with the same strength per unit mass, has a characteristic power law size distribution with a slope of 2.5. For a collisional population of gravity controlled (strengthless) bodies \citet{OBrien+Greenberg03} demonstrated that the expected size distribution has a shallower slope, 2.04.
Happily, the size distribution is also one of the more straightforward characteristics of the population to determine, as at least reasonable estimates of the size of bodies can be made with snap-shot observations. Time-series data allow better measurements of their sizes, as this removes uncertainties due to their rotational light curve. For Jupiter Family comets (JFCs) these observations are generally made when the comet is at a large distance from the Sun, and therefore more likely to be inactive, so a brightness measurement for the bare nucleus can be made. Converting the measured optical magnitude to the size of an object depends on its albedo and phase function. There are only 7 JFCs for which both of these are independently well measured (2P/Encke, 9P/Tempel~1, 10P/Tempel~2, 19P/Borrelly, 28P/Neujmin~1, 67P/Churyumov-Gerasimenko and 81P/Wild~2; three of which have a size measurement from resolved imaging by spacecraft [see Tables \ref{beta_table} and \ref{albedo_table} for references]). A few more comets have measurements of one or the other. 
Where these parameters are not known, we are forced to assume values for them; typically 4\% for the albedo and a linear phase function of $\beta = 0.035$ mag deg$^{-1}$. Previous measurements of the size distribution of JFCs have followed  similar assumptions. 

\begin{figure}
   \centering
   \includegraphics[angle=-90,width=0.45\textwidth]{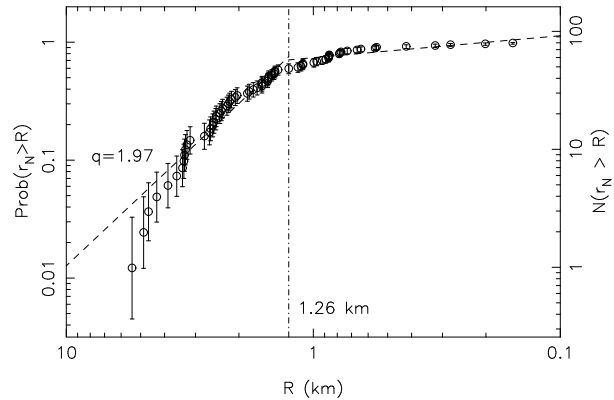} 
   \caption{Reference CSD with our data set and the usual assumptions. This shows the normal shape of the CSD, with a linear part and a cut-off. The error bars shown are calculated using the Kaplan-Meier statistic. Each fit within a MC run produces a CSD like this with a slightly different distribution of points.
   }
   \label{ref_CSD}
\end{figure}

The distribution is generally plotted (fig.~\ref{ref_CSD}) as a cumulative size distribution (CSD), expressed in terms
\begin{equation}
N(>r_{\rm N}) \propto r_{\rm N}^{-q}
\end{equation}
$N$ is the number of nuclei with radius, $r_{\rm N}$, larger than $r_{\rm N}$, and $q$ is the slope. A number of measurements of these distribution co-efficients have been made, generally based on snap-shot observations of a large number of nuclei. \citet{Lowry03} estimated $q = 1.6\pm0.1$ based on a sample of 33 comets, and \citet{Weissman+Lowry03} updated this to $q=1.59\pm0.03$, based on 41 JFCs with $r_{\rm N} \ge 1.4$ km. They chose to fit the size distribution to only those nuclei with $r_{\rm N} \ge 1.4$ km as the slope of the CSD is approximately constant above this radius, while below it there is a sharp cut-off. This break may imply a relative paucity of small nuclei compared with the expected number from a continuation of the power law from larger sizes; \citet{Meech04} and \citet{Fitzsimmons10} show that such a break is most likely real and not due to observational biases (as suggested by \citet{Lamy-chapter}) by modelling realistic observational surveys. A more recent size distribution from \citet{Weissman09} has a value of  $q=1.94\pm0.07$ based on 41 JFCs with $r_{\rm N} > 1.4$ km. Another estimate comes from \citet{FernandezJ99}, who used selected data in quality classes 1-3 (uncertainties on $m_V(1,1,0)$ up to $\pm 1$ mag.) from the catalogue presented by \citet{Tancredi00}, with cut-offs in both absolute magnitude and perihelion distance. The discrepancy between \citeauthor{FernandezJ99}'s estimate of $q=2.65\pm0.25$ and those of \citeauthor{Lowry03} can be explained by these cut-offs (leaving only 12 comets) and the large uncertainties on magnitudes in \citeauthor{Tancredi00}'s catalogue. \citet{Tancredi06} presented an updated catalogue and find $q=2.7\pm0.3$ for $r_{\rm N} \ge 1.5$ km. \citet{Meech04} estimate $q=1.45\pm0.05$ over the range $1 \le r_{\rm N} \le 10$ km, and a steeper $q=1.91\pm0.06$ in the range $2 \le r_{\rm N} \le 5$ km, showing the large dependence on the choice of size range. 
\citet{Hicks07} estimate $q=1.50\pm0.08$ from Near Earth Asteroid Tracking (NEAT) survey data, although this makes use of observations of active comets and a coma subtraction technique, so is not included with the other surveys of distant inactive comets in Table \ref{previous}. Finally, \citet{Lamy-chapter} collated the data from most of these catalogues, together with their own unpublished results, those from \citet{Licandro00b} and also from other papers on individual comets, to calculate $q=1.9\pm0.3$ for JFC nuclei with $r_{\rm N} \ge 1.6$ km. 

\begin{table}
\caption{Previous JFC size distribution estimates.}
\begin{center}
\begin{tabular}{l c c c}
Reference & $N_{\rm fit} / N_{\rm tot}$ & $r_{\rm N}$ range & $q$ \\
\hline
\citet{FernandezJ99} & 12/64 & $2.1 - 3.3$ & $2.7 \pm 0.3$ \\
\citet{Lowry03} & 16/19 & $1.4 - 3.6$ & $1.6 \pm 0.1$ \\
\citet{Weissman+Lowry03} & 41/54 & $1.4 - 15$ & $1.59 \pm 0.03$ \\
\citet{Meech04} & 38/48 & $1.0 - 10$ & $1.45 \pm 0.05$ \\
\citet{Meech04} & 21/48 & $2.0 - 5$ & $1.91 \pm 0.06$ \\
\citet{Lamy-chapter} & 29/65 & $1.6 - 15$ & $1.9 \pm 0.3$ \\
\citet{Weissman09} & 41/67 & $1.4 - 6.0$ & $1.94 \pm 0.07$ \\
\citet{Tancredi06} & 32/72 & $1.7 - 4.5$ & $2.7 \pm 0.3$ \\
\hline
\end{tabular}
\end{center}
\label{previous}
\end{table}%

These previous estimates are listed in Table \ref{previous}, where we list the number of comets included in the fit (and the total number considered in the survey), the range in sizes over which the authors fit the linear part of the CSD, and the resulting slope $q$. The uncertainty on $q$ is that quoted by each author, and is generally the formal uncertainty from a least squares (or similar) fit to the line, despite the fact that technically one cannot use such a fitting technique as the points in a cumulative distribution are not independent. These previous works also make no attempt to assess the uncertainty on $q$ contributed by the assumptions on phase function and albedo or the uncertainty from the photometry. In this paper we present an updated size distribution based on new photometry of distant JFC nuclei, using a censored data analysis technique to produce the CSD, and make a rigourous assessment of the uncertainty on $q$ due to all the unknown factors.

\section{Method}

\subsection{Censored data analysis}

The majority of previous measurements of the comet nucleus CSD follow a similar procedure. For each nucleus a best estimate of the intrinsic radius is arrived at either from the investigators own data or though consideration of several reported detections. The resulting radii are combined to form a CSD and a straight line is fit through the CSD using linear regression. Although this is relatively straightforward, there are intrinsic problems to this method. Aside from the incorrect use of slope uncertainties from linear regression techniques mentioned above, a more important drawback is that it ignores non-detections or  upper limits to nuclear radii. Not only is this is an important source of information, ignoring these non-detections can act to instill significant bias. For example, consider a single imaging survey of inactive nuclei with a well defined limiting magnitude for detection. Nearby comets might all be detected, but the fraction of comets observed would  decrease as their geocentric and heliocentric distances increased. The resulting CSD of nuclei would contain contain a progressively smaller fraction of the true population at smaller sizes, giving an observed CSD with a shallower power-law distribution than in reality. In the past, breaks in observed luminosity distributions have often been interpreted as due to the onset of such incompleteness. To our knowledge, the only previous study that has accounted for upper limits was that of \citet{Meech04}.

Many fields of astronomical research face a similar problem of censored data, where a sample contains both directly measured values and upper limits. The combination of observed and censored data was first tackled by \citet{Avni80} via a maximum likelihood method. Established statistical methods as applied to censored astronomical datasets were subsequently described by \citet{Schmidt85} and \citet{Feigelson+Nelson85}. These authors showed how the cumulative distribution function of a censored dataset can be constrained via the Kaplan-Meier estimator \citep{Kaplan+Meier58}.  This statistic normally estimates the probability of measuring a value for an observable less than or equal to a given value at the non-censored values of the dataset, or in our case for radii of $R$ km, 
$P(r_N \leq R)$. As long as the observed sample is randomly picked from the population, then this will be equivalent to estimating the relative number of comets with radii $\leq R$. Inverting  this function then gives the normal form of the CSD.

For cometary nuclei at large heliocentric distances the probability of non-detection will be larger as nucleus size decreases, hence at first sight the sample will be biased towards large nuclei. However, other effects act to randomise the censoring. First, comets are observed at various positions in their orbits, which when combined with different sized telescopes, imaging cameras and exposure times leads to a wide variation in effective detection limits.
As an example, a 600-second exposure at R-band with the 4.2m William Herschel Telescope can detect a point source at $m_R=24.5$ at a signal-to-noise of $\sim6$. Assuming an optical albedo of $0.04$, this allows direct detection of a  $r_N=0.8$km nucleus at opposition at $R_h=5.2$ AU, or the same nucleus at opposition at
$\Delta=R_h=4$AU but at a phase angle of $\simeq 20^\circ$. Second, when comets become active the shielding effect of their dust comae can result in censoring of even large cometary nuclei. As the strength and onset of activity appears to strongly vary from comet to comet, this further randomises whether an inert nucleus will be directly imaged or not. In this study we therefore assume that the observed cometary nuclei approximates a randomised sample of the JFC population.

The use of the Kaplan-Meier statistic is accurate as long as the quantity being measured and the censoring variable (here the nuclear radius and the limiting magnitude) are independent. The limiting magnitude is set by the instrumental setup, exposure times, and activity level of the comet, while the individual comets surveyed will normally depend on their orbital visibility at a particular date. Therefore this requirement should be true of most cometary observations where non-detection is due to large distance leading to a faint apparent magnitude. For nucleus measurements censored by activity, it will also hold. Although weak correlations of active area with orbital parameters have previously been suspected \citep{AHearn95}, there appears to be no link between nuclear radius and orbit \citep{Lamy-chapter}, and the orbital positions of comets are randomised by the dates of observation. Finally, an important aspect of the Kaplan-Meier statistic is that in principle it allows an estimate of the uncertainty of the luminosity function at each value where an uncensored measurement is made. One method of calculating this is given in equations (11) and (12) of \citet{Feigelson+Nelson85}, which we adopt in this work. We therefore plot the CSD in terms of $P(r_{\rm N} > R)$ with uncertainties at each point from this statistic, which is equivalent to $N(r_{\rm N} > R)$ and gives the slope $q$ we are interested in. \citet{Fitzsimmons10} give more details on this method and its application to comet size distributions.


\subsection{Fitting the CSD}

\begin{figure}
   \centering
   \includegraphics[angle=-90,width=0.45\textwidth]{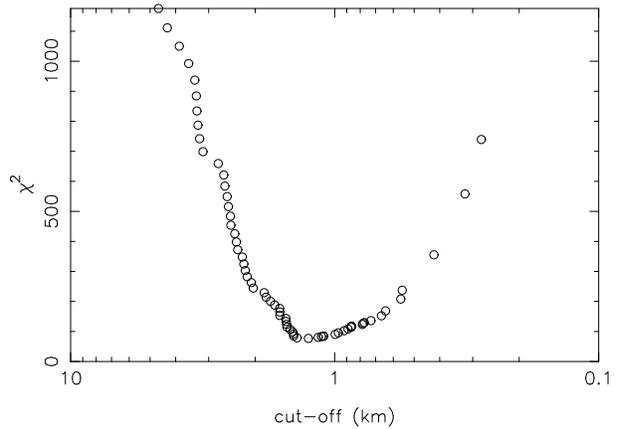} 
   \caption{Combined $\chi^2$ for two power law fits either side of a cut-off, as a function of the selected cut-off radius, for the reference CSD shown in fig. \ref{ref_CSD}. We select the cut-off radius with the minimum $\chi^2$ as the best description of the overall shape of the CSD, and therefore the corresponding $q$ value from the fit to the CSD at larger radii as the best slope for that particular CSD. In this case, the minimum is at a radius of 1.26 km.
   }
   \label{cutoff_chisq}
\end{figure}

We assume that the CSD is well described by two power laws (i.e. straight lines on a log-log plot), as we expect a break in the power law at small sizes. The power law describing the CSD at larger sizes is the one of primary interest, and is the one we refer to when discussing the slope $q$ elsewhere in this paper, as the data at smaller sizes suffers more from observational incompleteness. The choice of radius to define this break is important, as some of the variation of slopes found by the various papers listed in Table \ref{previous} can be explained by the different radius ranges over which the authors chose to fit the CSD. We attempt to independently choose the correct break radius by testing all possible radii in the range of the comet sizes. We fit two straight lines on either side of the break, and compute the $\chi^2$ statistic for the fit of this model to the CSD, for each test radius (fig. \ref{cutoff_chisq}). The minimum $\chi^2(r)$ then gives us the optimum break radius, which tends to fall around 1.2 km.

While strictly speaking a $\chi^2$ fit is not appropriate in this case, as a cumulative distribution does not have independent points, our goal is to find a simple description of the fit that gives a reasonable measure in an individual case. The formal uncertainty on each fit is not used, since we find the overall uncertainty from the standard deviation around the average of all fits. We investigated the results from a number of fitting techniques, to be sure that the result from the $\chi^2$ technique is representative. We performed these tests on our reference size distribution (fig. \ref{ref_CSD}; see section \ref{section:data}). The $\chi^2$ minimisation technique gives a value of $q=1.97$ for this data set.

One common approach to fitting a line to a cumulative distribution is to use a maximum-likelihood fit. Using the formulae given by \citet{Badenes2010} for the maximum likelihood power law index, we find $q=1.95\pm0.29$, a nearly identical result to the $\chi^2$ method.

An alternative, as used by \citet{Lamy-chapter}, is to test the observed distribution against various model distributions using a
Kolmogorov-Smirnov test. In this case the best fit is the one for which the probability that the samples are drawn from the same distribution is the highest. Applying this approach to our test size distribution returns a best fit value of $q = 2.03$ with a probability $P_{\rm KS}$ of 0.99, for the same cut off radius. This method has a significant draw back though, as the Kolmogorov-Smirnov test is not valid when the parameters of the model (in this case the cut-off radius and normalisation of the power law) are estimated from the data, which is necessary in this work.

A further disadvantage with both the maximum-likelihood and K-S techniques is that they fail to take advantage of the fact that the K-M statistics return uncertainties on each radius, which include information from upper limits and should be taken into account in an ideal fit. On the other hand, while the $\chi^2$ fit is not appropriate due to the fact the points are not independent, the use of K-M statistics mean that the error bars give proper weighting to each point. In conclusion, while none of these statistical methods are fully appropriate to these data, the fact that they all result in a similar value for $q$ gives us confidence that the result from $\chi^2$ is a reasonable estimate of the power-law exponent for the size distribution over this size range.


\subsection{Monte Carlo techniques}

The effective radius of the nucleus $r_{\rm N}$ is given by
\begin{equation}\label{rneqn}
A_R r^2_{\mathrm{N}} = 2.238 \times 10^{22} R^2_{\rm h} \Delta^2 10^{0.4(m_\odot - m_R + \beta\alpha)}
\end{equation}
where $A_R$ is the geometric albedo, $R_{\rm h}$, $\Delta$ and $\alpha$ are the heliocentric and geocentric distances (AU) and phase angle (degrees) at the time of observation, $\beta$ is the linear phase coefficient (mag deg$^{-1}$) and $m_R$ and $m_\odot=-27.09$ are the apparent magnitudes of the comet and the Sun, both in the $R$-band \citep{Russell16}. The orbital position parameters ($R_{\rm h}$, $\Delta$, $\alpha$) for a given observation and the magnitude of the Sun are known to a high level of precision, but the remaining parameters all contribute to the uncertainty on the reported radius. In addition, this equation converts a cross-sectional area (found as the area is directly proportional to the reflected flux) to an effective radius assuming a spherical nucleus, while the shape is also generally unknown (see section \ref{sec:shape}). We examine the effect of each of these sources of uncertainty by applying a Monte Carlo (MC) technique. For each comet observation in our data set (Table \ref{photometry_table}) we select a random value for each of the unknown parameters ($\beta$, $A_R$) from a distribution of possible values and a magnitude within the range $m_R \pm \sigma_R$, and use this to generate a radius. We correct this radius for the effects of the non-spherical shape of the nucleus by selecting an ellipsoidal shape from a suitable distribution of axial ratios. We then generate a CSD by fitting the resulting radii as described in the previous subsections. We repeat this procedure many times (generally, $N = 1000$), selecting a different random value for the parameters for each comet each time, which allows us to measure the effect of varying the parameters within our chosen distributions. We then measure the average value of $<q>$ found from the $N$ different CSDs; the standard deviation on this $\sigma(q)$ then incorporates the uncertainty on the varied parameters. In section \ref{variations} we vary each parameter in turn (while keeping the other unknown values fixed at default values) to assess the relative contribution to the total uncertainty of each parameter. Finally we allow all of the parameters to vary simultaneously to measure the overall uncertainty on $q$.

\section{Data set}\label{section:data}

\begin{table*}
\begin{minipage}{156mm}
\caption{Photometry database used for input}
\begin{center}
\begin{tabular}{l c c c r c c r l}
\hline
Comet & $R_{\rm h}$ & $\Delta$ & $\alpha$ & $m_R$ & $\sigma_R$ & $\Delta m_R$\footnote{%
The observed variation in magnitude, corresponding to a minimum axial ratio.} & $r_{\rm N}$\footnote{%
The radius (km) calculated for a \emph{fixed} $\beta=0.035$ mag deg$^{-1}$ and albedo of 4\%, for reference.} & Reference\\
\hline
2P/Encke\footnote{%
We include 2P/Encke, which is an ecliptic comet although not technically a JFC by the Tisserand parameter definition.
}  & 3.97 & 2.97 & 1.20 & 19.34 & 0.00 & 0.35 & 4.65 & \citet{FernandezY05} \\ 
4P/Faye  & 2.96 & 2.00 & 5.60 & 20.12 & 0.03 & -- & 1.75 & \citet{Lamy09} \\ 
6P/dArrest  & 2.83 & 2.19 & 18.19 & 20.74 & 0.07 & 0.18 & 1.69 & \citet{Lowry+Weissman03} \\ 
7P/Pons-Winnecke  & 4.69 & 4.31 & 11.60 & 22.46 & 0.02 & 0.30 & 2.24 & \citet{Snodgrass05} \\ 
9P/Tempel 1  & 3.52 & 4.03 & 13.30 & 21.30\footnote{%
From reported $V$-magnitude and $(V-R)=0.50\pm0.01$ from \citet{Li07}} %
& 0.04 & 0.50 & 2.72 & \citet{Lamy07} \\ 
10P/Tempel 2  & 3.99 & 3.16 & 9.00 & 19.66 & 0.05 & 0.35 & 4.86 & \citet{Jewitt+Luu89} \\ 
14P/Wolf  & 5.51 & 4.96 & 8.90 & 22.28 & 0.01 & 0.55 & 3.15 & \citet{Snodgrass05} \\ 
17P/Holmes  & 4.66 & 3.92 & 9.00 & 22.86 & 0.02 & 0.30 & 1.61 & \citet{Snodgrass06} \\ 
19P/Borrelly  & 1.40 & 0.62 & 38.00 & 16.77 & 0.05 & 1.00 & 2.03 & \citet{Lamy98} \\ 
21P/Giacobini-Zinner  & 3.80 & 3.21 & 13.60 & 21.89 & 0.04 & -- & 1.82 & \citet{Pittichova08} \\ 
22P/Kopff  & 4.49 & 3.80 & 9.93 & 21.23 & 0.10 & 0.58 & 3.25 & \citet{Lowry+Weissman03} \\ 
26P/Grigg-Skjellerup  & 3.82 & 2.85 & 5.00 & 21.60 & 0.07 & -- & 1.61 & \citet{Boehnhardt99} \\ 
28P/Neujmin 1  & 7.66 & 6.67 & 1.53 & 20.74 & 0.02 & 0.45 & 10.65 & \citet{Delahodde01} \\ 
29P/Schwassmann-Wachmann 1 & 5.89 & 4.97 & 4.62 & $\ge$18.00 & 0.01 & -- &$\le$ 22.63 & \citet{Meech93} \\ 
31P/Schwassmann-Wachmann 2  & 4.58 & 3.69 & 6.35 & 21.05 & 0.06 & 0.50 & 3.29 & \citet{Luu+Jewitt92} \\ 
36P/Whipple  & 4.78 & 3.79 & 1.00 & 21.57 & 0.01 & 0.70 & 2.55 & \citet{Snodgrass08} \\ 
37P/Forbes  & 2.27 & 1.39 & 14.80 & 20.83 & 0.04 & -- & 0.78 & \citet{Lamy09} \\ 
40P/Vaisala 1  & 4.58 & 3.68 & 6.40 & $\ge$22.08 & 0.02 & -- &$\le$ 2.05 & \citet{Snodgrass08} \\ 
43P/Wolf-Harrington  & 4.43 & 3.44 & 3.80 & $\ge$21.45 & 0.06 & -- &$\le$ 2.38 & \citet{Lowry+Fitzsimmons05} \\ 
44P/Reinmuth 2  & 4.51 & 3.84 & 10.10 & 22.49 & 0.06 & -- & 1.85 & \citet{Snodgrass08} \\ 
45P/Honda-Mrkos-Pajdusakova  & 5.14 & 4.17 & 3.69 & 23.27 & 0.86 & -- & 1.44 & \citet{Lowry03} \\ 
46P/Wirtanen  & 4.98 & 4.00 & 3.50 & 25.15 & 0.15 & 0.38 & 0.56 &  \citet{Boehnhardt02} \\ 
47P/Ashbrook-Jackson  & 5.42 & 4.48 & 3.50 & 21.68 & 0.01 & 0.45 & 3.39 & \citet{Snodgrass06} \\ 
48P/Johnson  & 3.87 & 2.94 & 6.10 & 21.11 & 0.02 & 0.32 & 2.14 &  \citet{Jewitt+Sheppard04} \\ 
49P/Arend-Rigaux  & 3.34 & 2.78 & 16.01 & 19.51 & 0.05 & -- & 4.30 & \citet{Lowry03} \\ 
50P/Arend  & 2.37 & 1.47 & 11.80 & 20.58 & 0.04 & -- & 0.92 & \citet{Lamy09} \\ 
51P/Harrington & 5.30 & 5.73 & 9.22 & $\ge$23.50 & 0.00 & -- &$\le$ 2.01 & \citet{Lowry+Fitzsimmons01} \\ 
52P/Harrington-Abell  & 2.83 & 1.86 & 5.50 & 20.30 & 0.20 & -- & 1.43 & \citet{Licandro00b} \\ 
53P/Van Biesbroeck  & 8.31 & 7.34 & 1.38 & 23.65 & 0.06 & -- & 3.32 & \citet{Meech04} \\ 
54P/de Vico-Swift & 5.39 & 5.70 & 9.68 & $\ge$23.20 & 0.00 & -- &$\le$ 2.35 & \citet{Lowry+Fitzsimmons01} \\ 
56P/Slaughter-Burnham  & 7.42 & 7.58 & 7.41 & 25.37 & 0.12 & -- & 1.53 & \citet{Meech04} \\ 
57P/duToit-Neujmin-Delporte & 5.10 & 4.35 & 7.82 & $\ge$24.00 & 0.00 & -- &$\le$ 1.14 & \citet{Lowry+Fitzsimmons01} \\ 
59P/Kearns-Kwee  & 2.52 & 1.54 & 3.30 & 20.86 & 0.05 & -- & 0.79 & \citet{Lamy09} \\ 
61P/Shajn-Schaldach  & 4.39 & 3.40 & 3.14 & 23.27 & 0.86 & -- & 1.00 & \citet{Lowry03} \\ 
63P/Wild 1  & 2.27 & 1.30 & 9.20 & 19.17 & 0.02 & -- & 1.43 & \citet{Lamy09} \\ 
64P/Swift-Gehrels  & 3.63 & 2.79 & 9.50 & 21.60\footnote{%
From reported $V$-magnitude and assumed $(V-R)=0.50$.}%
 & 0.10 & -- & 1.61 & \citet{Licandro00b} \\ 
65P/Gunn & 4.43 & 3.55 & 6.36 & $\ge$17.74 & 0.04 & -- &$\le$ 14.09 & \citet{Lowry+Fitzsimmons01} \\ 
67P/Churyumov-Gerasimenko  & 5.60 & 4.60 & 0.50 & 22.46 & 0.01 & 0.40 & 2.39 & \citet{Tubiana08} \\ 
69P/Taylor  & 4.89 & 3.99 & 6.10 & 21.00 & 0.40 & -- & 3.88 & \citet{Lowry99} \\ 
70P/Kojima  & 4.84 & 4.26 & 10.60 & 22.53 & 0.15 & -- & 2.18 & \citet{Snodgrass08} \\ 
71P/Clark  & 2.71 & 1.76 & 7.50 & 21.83 & 0.07 & -- & 0.66 & \citet{Lamy09} \\ 
73P/Schwassmann-Wachmann 3  & 3.03 & 2.35 & 15.00 & 21.57 & 0.05 & -- & 1.26 &  \citet{Boehnhardt99} \\ 
74P/Smirnova-Chernykh  & 3.55 & 2.58 & 4.20 & 20.52 & 0.10 & 0.14 & 2.21 &\citet{Lamy-chapter}\\ 
75P/Kohoutek  & 3.48 & 2.73 & 12.36 & 21.76 & 0.05 & -- & 1.47 & \citet{Weissman08} \\ 
76P/West-Kohoutek-Ikemura  & 3.09 & 2.26 & 12.10 & 24.39 & 0.10 & 0.42 & 0.32 &\citet{Lamy-chapter}\\ 
78P/Gehrels 2  & 5.46 & 4.49 & 3.54 & 23.44 & 0.19 & -- & 1.52 & \citet{Lowry+Weissman03} \\ 
79P/duToit-Hartley  & 4.74 & 4.29 & 11.50 & 23.30 & 0.40 & -- & 1.53 & \citet{Lowry99} \\ 
81P/Wild 2  & 5.29 & 4.31 & 2.06 & 22.13 & 0.21 & -- & 2.52 & \citet{Weissman08} \\ 
82P/Gehrels 3  & 3.73 & 2.75 & 0.60 & 23.04 & 0.10 & 0.51 & 0.73 &\citet{Lamy-chapter}\\ 
84P/Giclas  & 2.21 & 1.37 & 16.90 & 20.59 & 0.04 & -- & 0.86 & \citet{Lamy09} \\ 
86P/Wild 3  & 4.95 & 4.10 & 6.23 & 25.00 & 0.10 & -- & 0.64 & \citet{Meech04} \\ 
87P/Bus  & 2.45 & 1.43 & 2.60 & 22.87 & 0.10 & 0.86 & 0.28 &\citet{Lamy-chapter}\\ 
92P/Sanguin  & 4.46 & 3.58 & 6.30 & 21.94 & 0.01 & 0.60 & 2.07 & \citet{Snodgrass05} \\ 
94P/Russell 4  & 4.14 & 3.18 & 5.30 & $\ge$20.97 & 0.01 & -- &$\le$ 2.62 & \citet{Snodgrass08} \\ 
97P/Metcalf-Brewington  & 4.76 & 4.11 & 10.11 & 22.23 & 0.40 & -- & 2.36 & \citet{Lowry03} \\ 
98P/Takamizawa  & 3.28 & 2.53 & 12.70 & 24.20 & 0.30 & -- & 0.42 & \citet{MazzottaEpifani08} \\ 
\hline
\end{tabular}
\end{center}
\label{photometry_table}
\end{minipage}
\end{table*}%
\setcounter{table}{1}
\begin{table*}
\begin{minipage}{156mm}
\caption{Photometry database used for input (continued)}
\begin{center}
\begin{tabular}{l c c c r c c r l}
\hline
Comet & $R_{\rm h}$ & $\Delta$ & $\alpha$ & $m_R$ & $\sigma_R$ & $\Delta m_R$ & $r_{\rm N}$ & Reference\\
\hline
100P/Hartley 1  & 4.84 & 4.06 & 8.20 & 22.06 & 0.06 & -- & 2.48 & \citet{Weissman08} \\ 
103P/Hartley 2  & 5.60 & 4.70 & 4.50 & $\ge$24.50 & 0.04 & -- &$\le$ 1.02 & \citet{Snodgrass10} \\ 
104P/Kowal 2  & 3.94 & 3.31 & 12.66 & 23.05 & 0.93 & -- & 1.12 & \citet{Lowry03} \\ 
106P/Schuster  & 1.67 & 0.78 & 23.00 & 18.91 & 0.03 & -- & 0.89 & \citet{Lamy09} \\ 
110P/Hartley 3  & 3.60 & 2.66 & 6.58 & 20.58 & 0.10 & -- & 2.33 & \citet{Weissman08} \\ 
111P/Helin-Roman-Crockett  & 3.47 & 2.53 & 5.60 & 21.48 & 0.05 & -- & 1.39 & \citet{MazzottaEpifani07} \\ 
112P/Urata-Niijima  & 2.30 & 1.50 & 19.20 & 20.96 & 0.04 & -- & 0.86 & \citet{Lamy09} \\ 
114P/Wiseman-Skiff  & 3.75 & 2.95 & 10.90 & 22.95 & 0.11 & -- & 0.97 & \citet{Snodgrass08} \\ 
115P/Maury  & 6.38 & 5.43 & 2.56 & 24.86 & 0.07 & -- & 1.10 & \citet{Meech04} \\ 
116P/Wild 4 & 4.72 & 3.82 & 6.00 & $\ge$21.36 & 0.06 & -- &$\le$ 3.03 & \citet{Weissman08} \\ 
117P/Helin-Roman-Alu 1  & 3.29 & 2.56 & 13.50 & $\ge$16.31 & 0.03 & -- &$\le$ 16.36 & \citet{MazzottaEpifani08} \\ 
118P/Shoemaker-Levy 4  & 4.71 & 3.72 & 3.50 & 21.54 & 0.20 & -- & 2.61 & \citet{Lowry03} \\ 
120P/Mueller 1  & 3.89 & 3.05 & 8.80 & 23.52 & 0.13 & -- & 0.77 & \citet{Snodgrass08} \\ 
121P/Shoemaker-Holt 2  & 3.92 & 3.43 & 13.50 & 20.77 & 0.01 & 0.15 & 3.34 & \citet{Snodgrass08} \\ 
128P/Shoemaker-Holt 1  & 4.99 & 4.16 & 7.29 & 22.02 & 0.09 & -- & 2.63 & \citet{Lowry+Weissman03} \\ 
131P/Mueller 2  & 3.48 & 2.65 & 10.10 & 22.77 & 0.09 & -- & 0.87 & \citet{Snodgrass08} \\ 
136P/Mueller 3  & 4.83 & 4.04 & 8.00 & 23.70 & 0.20 & -- & 1.15 & \citet{MazzottaEpifani08} \\ 
137P/Shoemaker-Levy 2  & 6.95 & 6.17 & 5.30 & 22.86 & 0.03 & -- & 3.57 & \citet{Snodgrass06} \\ 
143P/Kowal-Mrkos  & 3.40 & 2.47 & 8.20 & 18.52 & 0.02 & -- & 5.42 & \citet{Jewitt02} \\ 
147P/Kushida-Muramatsu  & 2.83 & 2.30 & 18.70 & 25.48 & 0.10 & 0.46 & 0.20 &\citet{Lamy-chapter}\\ 
160P/LINEAR & 3.98 & 3.41 & 12.50 & 23.69 & 0.17 & -- & 0.87 & \citet{Snodgrass08} \\ 
179P/Jedicke & 5.52 & 4.71 & 6.90 & 22.63 & 0.07 & -- & 2.47 & \citet{Snodgrass08} \\ 
243P/NEAT & 3.97 & 3.68 & 14.30 & 22.70 & 0.30 & -- & 1.52 & \citet{MazzottaEpifani08} \\ 
D/1819 W1 (Blanpain) & 1.64 & 0.72 & 20.70 & 22.40 & 0.10 & -- & 0.16 & \citet{Jewitt06} \\ 
P/2001 H5 (NEAT) & 4.61 & 3.62 & 3.13 & 23.29 & 0.11 & -- & 1.10 & \citet{Lowry+Fitzsimmons05} \\ 
P/2002 T5 (LINEAR) & 5.24 & 4.46 & 7.50 & $\ge$19.10 & 0.30 & -- &$\le$ 11.40 & \citet{MazzottaEpifani08} \\ 
P/2003 S1 (NEAT) & 3.54 & 3.05 & 15.20 & $\ge$21.37 & 0.03 & -- &$\le$ 2.10 & \citet{MazzottaEpifani08} \\ 
P/2004 DO29 (Spacewatch-LINEAR) & 4.22 & 3.33 & 7.00 & $\ge$20.13 & 0.03 & -- &$\le$ 4.23 & \citet{MazzottaEpifani08} \\ 
P/2004 H2 (Larsen) & 3.71 & 3.04 & 13.10 & $\ge$21.59 & 0.01 & -- &$\le$ 1.91 & \citet{Snodgrass08} \\ 
P/2004 H3 (Larsen) & 3.71 & 2.97 & 12.00 & 24.19 & 0.21 & -- & 0.55 & \citet{Snodgrass08} \\ 
\hline
\end{tabular}
\end{center}
\end{minipage}
\end{table*}%


The data used for this investigation are selected from all JFC nucleus photometry found in the literature (only properly calibrated professional data sets), including our own previously published work. The majority come from the surveys listed in the introduction that made their own estimates of the CSD, which were all collated in the review by \citet{Lamy-chapter}. We also include the many observations published in the six years since the publication of that work, which have added new comets to the database and improved the accuracy of the radius measurements of others (in many cases measuring light curves where previously only snap-shot measurements had been taken). We include the surveys of \citet{Lowry+Fitzsimmons05}, \citet{Snodgrass05,Snodgrass06,Snodgrass08}, \citet{MazzottaEpifani07,MazzottaEpifani08} and \citet{Weissman08}, and also papers on individual comets from \citet{FernandezY05} [2P/Encke], 
\citet{Jewitt06} [D/1819 W1 (Blanpain)], \citet{Snodgrass10} [103P/Hartley 2], \citet{Tubiana08} [67P/Churyumov-Gerasimenko] and \citet{Pittichova08} [21P/Giacobini-Zinner]. The references for each comet are given in Table \ref{photometry_table}.

To apply our MC method, it is necessary that we choose one observation at a given $R_{\rm h}$, $\Delta$ and $\alpha$ for each comet, and use equation \ref{rneqn} to generate the corresponding radius for a chosen set of parameters. For some comets there have been multiple observations at different epochs, and we therefore require suitable criteria to select the `best' observation. We must do this rather than attempting to take an average value of the radius for the comet from separate observations (as was done by \citet{Lamy-chapter}, for example) as such a combination would require an assumed phase function to correct for the different orbital positions, and the effect of the unknown phase function is one of the parameters we investigate.

The criteria we use are:
\begin{enumerate}
\item Is there a detection of the inactive nucleus? If not, we use the strongest upper limit, and flag the value as a limit (except in the cases where space based observations allowed an accurate coma subtraction). 
\item Is there a light curve? Where there is a direct nucleus detection we prefer light curves over snap-shots, as this removes some of the uncertainty due to the unknown shape.
\item Finally, where we are left with a choice between snap-shot observations, we use the result with the smallest photometric uncertainty and smallest phase angle to minimise uncertainty.
\end{enumerate}

The final list of selected observations is given in Table \ref{photometry_table}.  The table also includes the radii calculated under the standard assumptions ($\beta=0.035$ mag deg$^{-1}$ and $A_R=0.04$ for all comets; photometric values are precise and correspond directly to the size of a spherical nucleus). This is for reference only as the radius found for each comet varies in each MC run, over a range set by the variable parameters, throughout the rest of the paper. The data set contains 86 comets. 71 are nucleus detections, 15 are upper limits from either active or non-detected comets. 21 have light curves, and another 2 have a constraint on a magnitude range but not a full light curve. 12 have known $\beta$ (although these are often very rough estimates), 11 have known albedo and 7 have both of these parameters measured.

For reference, fitting a CSD to this data set without varying any parameters (i.e. following the procedure applied to previous works; taking the reference radii based on the standard assumptions listed in Table \ref{photometry_table}) gives $q=1.97$, with a cut off at 1.26 km (fig. \ref{ref_CSD}).  For comparison, using the measured phase functions and albedos for the comets where these values are known (and using the assumed values given above for other comets) gives $q=2.00$, and the same cut off. There are relatively few comets with known $\beta$ and/or $A_R$ (see Tables \ref{beta_table} and \ref{albedo_table}), and these tend to be the larger comets which have a weak affect on the CSD, so the difference is not large. 

\section{Variation of parameters}\label{variations}

\subsection{Photometric uncertainty} 

The first source of uncertainty in the radius, and the most straight forward to account for, is the uncertainty $\sigma_R$ in the measured magnitude of the comet $m_R$. This includes the Poisson noise from the CCD photometry, uncertainty from the standard star calibration, etc. The true value can fall anywhere within the Gaussian distribution centred on the reported value, with the standard deviation set by the reported error bar. In most cases $\sigma_R \le 0.1$ mag, and this is not the most significant uncertainty on $r_{\rm N}$. The result of a MC run allowing the comet magnitudes to vary within the Gaussian distributions defined by the reported $m_R$, $\sigma_R$ for each, and holding all other parameters fixed at the default values given above, gives $q = 1.93 \pm 0.10$ and an average cut-off of 1.2 km.


\subsection{Phase function}

\begin{table}
\caption{Phase function measurements for JFCs.}
\begin{center}
\begin{tabular}{l c l}
\hline
Comet & $\beta$ & Reference.\\
\hline
2P   & 0.060 $\pm$ 0.005 & \citet{FernandezY00} \\ 
~" & 0.049 $\pm$ 0.004 & \citet{Boehnhardt08} \\
~" & {\it 0.053 $\pm$ 0.003} & {\it Weighted mean} \\
9P   & 0.046 $\pm$ 0.007 & \citet{Li07} \\ 
10P  & 0.037 $\pm$ 0.004 & \citet{Sekanina91} \\ 
19P  & 0.043 $\pm$ 0.009 & \citet{Li07b} \\ 
28P  & 0.025 $\pm$ 0.006 & \citet{Delahodde01} \\ 
36P  & 0.060 $\pm$ 0.019 & \citet{Snodgrass08} \\ 
45P & $\sim$ 0.06 & \citet{Lamy-chapter} \\
47P  & 0.083 $\pm$ 0.006 & \citet{Snodgrass08} \\ 
48P  & 0.059 $\pm$ 0.002 & \citet{Jewitt+Sheppard04} \\ 
67P  & 0.076 $\pm$ 0.003 & \citet{Tubiana08} \\ 
81P & 0.0513 $\pm$ 0.0002 & \citet{Li09} \\
143P  & 0.043 $\pm$ 0.001 & \citet{Jewitt03} \\ 
\hline
\end{tabular}
\end{center}
\label{beta_table}
\end{table}%

\begin{figure}
   \centering
   \includegraphics[angle=-90,width=0.5\textwidth]{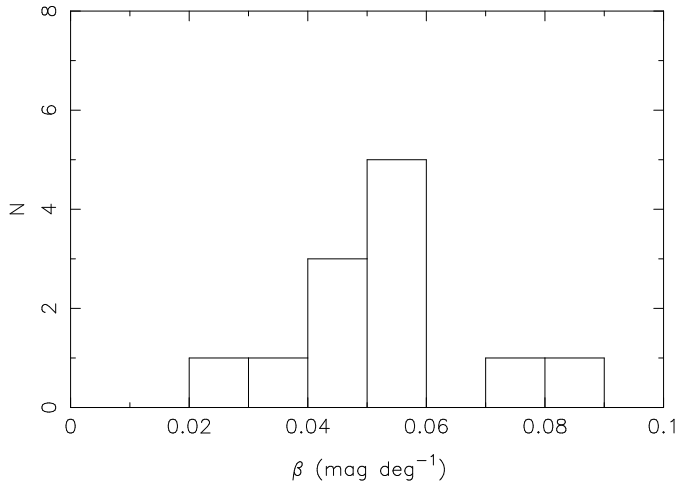} 
   \caption{Distribution of measured phase functions for JFCs.}
   \label{beta_fig}
\end{figure}

For all previous size distributions, a constant Solar phase function has been assumed for all comets. The work presented in this paper was motivated in part by a simple demonstration that changing the assumed value changed the resulting value of $q$: increasing the assumed $\beta$ increases $q$ \citep{Snodgrass-thesis}. To investigate the more realistic situation of $\beta$ varying between comets within some distribution we developed the MC technique presented here. We recreate the earlier result when choosing the simplest distribution; a constant value for all comets.

To choose an appropriate distribution we briefly review the measured values. Phase functions for comets are difficult to measure, since it is necessary to obtain light curves at multiple epochs to remove rotational variations from the photometry and leave the variation due to changing phase angle. Even relatively well measured phase curves, such as that for 28P/Neujmin 1 \citep{Delahodde01}, are still quite uncertain, and rough estimates from only a few epochs are very approximate even if formally good fits to the data \citep[e.g.][]{Snodgrass08}. The best results come from spacecraft data, although we note that spacecraft may measure phase functions to high $\alpha$, while ground based measurements generally only cover $\alpha \la 15\degr$. We list all JFCs with estimates in Table \ref{beta_table}, and plot the distribution of values in fig.~\ref{beta_fig}. It is noticeable that the measured phase functions are generally steeper than the normally assumed values of $\beta=0.035$ or 0.04 mag deg$^{-1}$; the mean value is 0.053 $\pm$ 0.016. This is due to a number of recent measurements of steep phase functions for JFCs, while the larger (brighter, easier to observe) nuclei observed previously all had shallower slopes. We tested for any correlation between size of nucleus and phase function, but none exists. The $\beta$ values are approximately normally distributed around the mean, so we use a Gaussian distribution for input into the MC simulations.

A MC run with a distribution matching this observed one gives $q= 1.95 \pm 0.07$. Increasing the mean $\beta$ increases $q$, and decreasing it decreases $q$, as expected. Changing the width of the distribution also affects the CSD slope, a narrower distribution ($\beta = 0.053 \pm 0.005$) gives a steeper slope and a smaller uncertainty, $q= 1.99 \pm 0.02$, while a wider distribution ($\sigma = 0.025$) gives a shallower slope and larger uncertainty ($q=1.91 \pm 0.11$).


\subsection{Albedo}

\begin{table}
\caption{Albedo measurements for JFCs.}
\begin{center}
\begin{tabular}{l c l}
\hline
Comet & $A_R$ & Reference\\
\hline
2P  & 0.050 $\pm$ 0.030& \citet{FernandezY00} \\ 
9P  & 0.064 $\pm$ 0.013 & \citet{Li07} \\
~" & 0.046 $\pm$ 0.015 & \citet{Lisse05}  \\
~" & 0.072 $\pm$ 0.016 & \citet{FernandezY03}  \\
~" & {\it 0.061 $\pm$ 0.008} & {\it Weighted mean}   \\
10P  & 0.030 $\pm$ 0.012 &\citet{Ahearn89}  \\ 
19P  & 0.029 $\pm$ 0.006  & \citet{Buratti04}  \\ 
~" & 0.072 $\pm$ 0.020   &\citet{Li07b}  \\
~" &  {\it 0.033 $\pm$ 0.006}   & {\it Weighted mean} \\
22P  & 0.048 $\pm$ 0.010 & \citet{Lamy02}  \\ 
28P  & 0.03 $\pm$ 0.01 &  \citet{Jewitt+Meech88}   \\ 
49P  & 0.045 $\pm$  0.019 & \citet{Campins95} \\ 
67P & 0.054 $\pm$ 0.006 & \citet{Kelley09} \\
81P  & 0.064 $\pm$ 0.010 & \citet{Li09}  \\ 
103P & 0.028 $\pm$ 0.009$^a$  & \citet{Lisse09}  \\
162P & 0.037 $\pm$ 0.014  & \citet{FernandezY06} \\
\hline
\end{tabular}
\end{center}
$^a$ Uncertain due to difficulties with the optical photometry, see \citet{Snodgrass10}
\label{albedo_table}
\end{table}%

\begin{figure}
   \centering
   \includegraphics[angle=-90,width=0.45\textwidth]{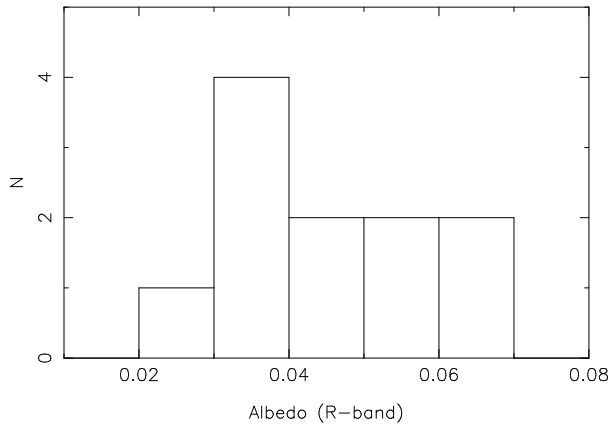} 
   \caption{Distribution of observed $R$-band albedos.}
   \label{fig:albedo_dist}
\end{figure}

The conversion from a reflected flux to a radius requires knowledge of the albedo, which generally must be assumed for comets. Under the normal assumption that all comets have the same albedo the CSD slope is directly proportional to the slope of the `luminosity' (reflected flux) distribution ($q_{\rm S} = 5 q_{\rm L}$, \citet{Irwin95}), and so a change of assumed albedo will not affect $q$. However, this does not hold for a distribution of albedo values that vary between comets, so we use our MC technique to assess the affect on the CSD of a variable albedo.

The distribution of albedos for JFC nuclei is poorly constrained. Only when the radius can be determined independently of the flux can the albedo be found, and this has been done for very few comets (generally by simultaneous observation in the optical and thermal infrared, where a thermal model of the nucleus returns the size for comparison with the reflected optical flux, or from spacecraft observations). We list all published JFC albedos in Table \ref{albedo_table}, converting albedos to $R$-band where necessary, and plot the distribution of $R$-band values in fig.~\ref{fig:albedo_dist}. We note that the values obtained by \citet{Li07,Li07b,Li09}, from modelling of disk resolved photometry from spacecraft data, are higher than all but one of the other measurements. This suggests some underlying problem in comparing results from this technique with radiometric methods, but we take them at face value for the purposes of this study. The mean value is $0.044 \pm 0.013$, close to the usually assumed 4\%. The small number of known values means that the shape of the distribution if not well defined; we choose to use a Gaussian distribution as this seems reasonable.

For a Gaussian with the observed mean and standard deviation, the CSD has $q=1.85 \pm 0.15$. For a lower average albedo (2\%) we find a shallow slope, $q =1.64 \pm 0.21$, with a larger uncertainty as some objects in such a distribution will occasionally have $A_R \approx 0$, giving very large sizes and therefore a large spread in slopes. A higher average albedo (6\%) gives $q= 1.89 \pm 0.12$, a steeper and less uncertain slope. A narrower distribution ($A_R = 0.04 \pm 0.005$) gives $q = 1.93 \pm 0.09$, while a broad distribution ($A_R = 0.04 \pm 0.02$) gives $q = 1.72 \pm 0.19$.


\subsection{Correction for shape}\label{sec:shape}

\subsubsection{Relating true effective radii and snap-shot radii}

Comets are generally found to be non-spherical, and the varying cross-sectional area of a rotating elongated body reflects a changing amount of light, which would imply a different effective radius if a `snap-shot' measurement was made at different rotational phase. Nuclei are often described as tri-axial ellipsoids, with semi-major axes $a \ge b = c$, as this is the simplest non-spherical body. Except for comets that have been visited by spacecraft, there generally is not sufficient data to make a more complex model of the shape. When producing a size distribution of these bodies, we are interested in the effective radius of a sphere with the same volume, given by $r_{\rm N} = \sqrt[3]{ab^2}$, but what we measure using equation~\ref{rneqn} is the radius corresponding to the projected cross sectional area at the time of observation. The projected area of a tri-axial ellipsoid, at rotational phase $\phi$ and with the pole orientated at $\epsilon$ to the line of sight, is given by \citep{Lamy-chapter}:
\begin{equation}\label{area_eqn}
S(\phi,\epsilon)=\pi a b^2 \sqrt{ \left(\frac{\sin^2{\phi}}{a^2} + \frac{\cos^2{\phi}}{b^2}\right)\sin^2{\epsilon} + \frac{\cos^2{\epsilon}}{b^2} }.
\end{equation}
The radius found from a snap-shot observation $r_{\rm ss}$ is the one corresponding to this area. Substituting $S(\phi,\epsilon) = \pi r_{\rm ss}^2$ and $ab^2 = r_{\rm N}^3$ into equation \ref{area_eqn} we find that the true effective radius is related to the snap-shot radius by:
\begin{equation}\label{snapshot_correction}
r_{\rm N} = r_{\rm ss} \left(\frac{a}{b}\right)^{-\frac{1}{6}} \left[ \left(\frac{\sin^2{\phi}}{\frac{a}{b}^2} + \cos^2{\phi}\right)\sin^2{\epsilon} + \cos^2{\epsilon} \right]^{-\frac{1}{4}}
\end{equation}
This equation can be used to find the \emph{average} relation between $r_{\rm N}$ and $r_{\rm ss}$ by integrating over $0 \le \phi \le 180\degr$ and $0 \le \epsilon \le 90\degr$, which gives a correspondence that is a weak function of $a/b$:
\begin{equation}
r_{\rm N} = r_{\rm ss} \left(\frac{a}{b}\right)^{-\frac{1}{6}} \sqrt{2}\left[3 + \left(\frac{a}{b}\right)^{-2}\right]^{-\frac{1}{4}}
\end{equation}
For an $a/b = 1.5$ this gives $r_{\rm N} = 0.97 r_{\rm ss}$, and for a more extreme $a/b = 3$ we find $r_{\rm N} = 0.89 r_{\rm ss}$. This implies that snap-shots will, \emph{on average}, give a very reasonable measurement of the effective radius, with the true radius being between 90-100\% of the reported radius from a single snap-shot.

For individual measurements this does not hold; sometimes an observation will happen to fall at light curve minimum, and sometimes it will fall at a maximum. Instead of applying a statistical correction to the whole database of snap-shots under the assumption that they will average out, we use MC simulations to generate the `true' radius of the nucleus from a snap-shot observation using equation \ref{snapshot_correction}. We do this by selecting a random shape ($a/b$) and orientation ($\phi$, $\epsilon$) for the nucleus, together with the radius $r_{\rm ss}$ from the snap-shot magnitude (and all other assumed parameters) given by equation~\ref{rneqn}. Of course the `true' radius generated this way may not match reality for any individual case, but over the $N$ MC runs this will generate a distribution of the possible effective radii that match the snap-shot photometry. This method has the advantage that it doesn't simply apply a correction for snap-shots that shift the radii up or down uniformly, but produces a family of CSDs describing how the overall distribution is affected by the possible variation in individual comet radii due to shape, allowing us to measure the uncertainty introduced.

The MC method requires that we choose appropriate distributions for the random parameters. For $\phi$ and $\epsilon$ this is trivial; any value in the ranges $0 \le \phi \le 180\degr$ and $0 \le \epsilon \le 90\degr$ respectively is equally likely as the rotational phase and pole orientation is entirely random for comets, as torques from jets can change the spin state in any direction on a short timescale. When a light curve is measured, one obtains the magnitude averaged over rotational phase, but $\epsilon$ is generally unknown. In this case we follow the same technique, but set $\phi=0$ and only randomise $\epsilon$.  The choice of axial ratio is more difficult, as we need to understand the distribution of nucleus shapes. In the next subsection we discuss the current knowledge of this distribution.

\subsubsection{Nucleus shapes} 

\begin{figure} 
   \centering
    \includegraphics[angle=-90,width=0.4\textwidth]{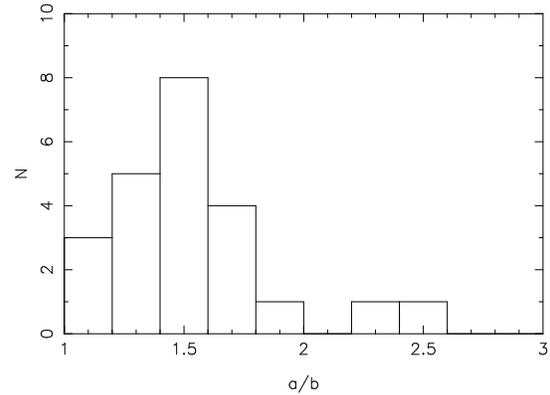}
   \caption{Distribution of observed {\it minimum} axial ratios for cometary nuclei, from the magnitude ranges listed in table \ref{photometry_table}.}
   \label{fig:abd_obs}
\end{figure}

Some constraint can be placed on the shape of an individual nucleus from its light curve, as a \emph{minimum} value of $a/b$ can be found from the amplitude of the light curve $\Delta m$:
\begin{equation}\label{delta_m_eqn}
\left(\frac{a}{b}\right)_{\rm min} = \frac{\pi ab}{\pi b^2} = \frac{{\rm Area}_{\rm max}}{{\rm Area}_{\rm min}} = \frac{{\rm Flux}_{\rm max}}{{\rm Flux}_{\rm min}} = 10^{0.4\Delta m}
\end{equation}
The distribution of minimum axial ratios from the light curves available in our catalogue is shown in figure \ref{fig:abd_obs}; it is peaked at $a/b \sim 1.5$ and can be reasonably well described as having a decrease in $N$ as $a/b$ increases, with a lack of bodies with $a/b\approx1$.
However, this describes only the observed distribution of minimum axial ratios, not the distribution of actual shapes. 

\begin{figure} 
   \centering
 \begin{tabular}{cc}
   \includegraphics[angle=-90,width=0.22\textwidth]{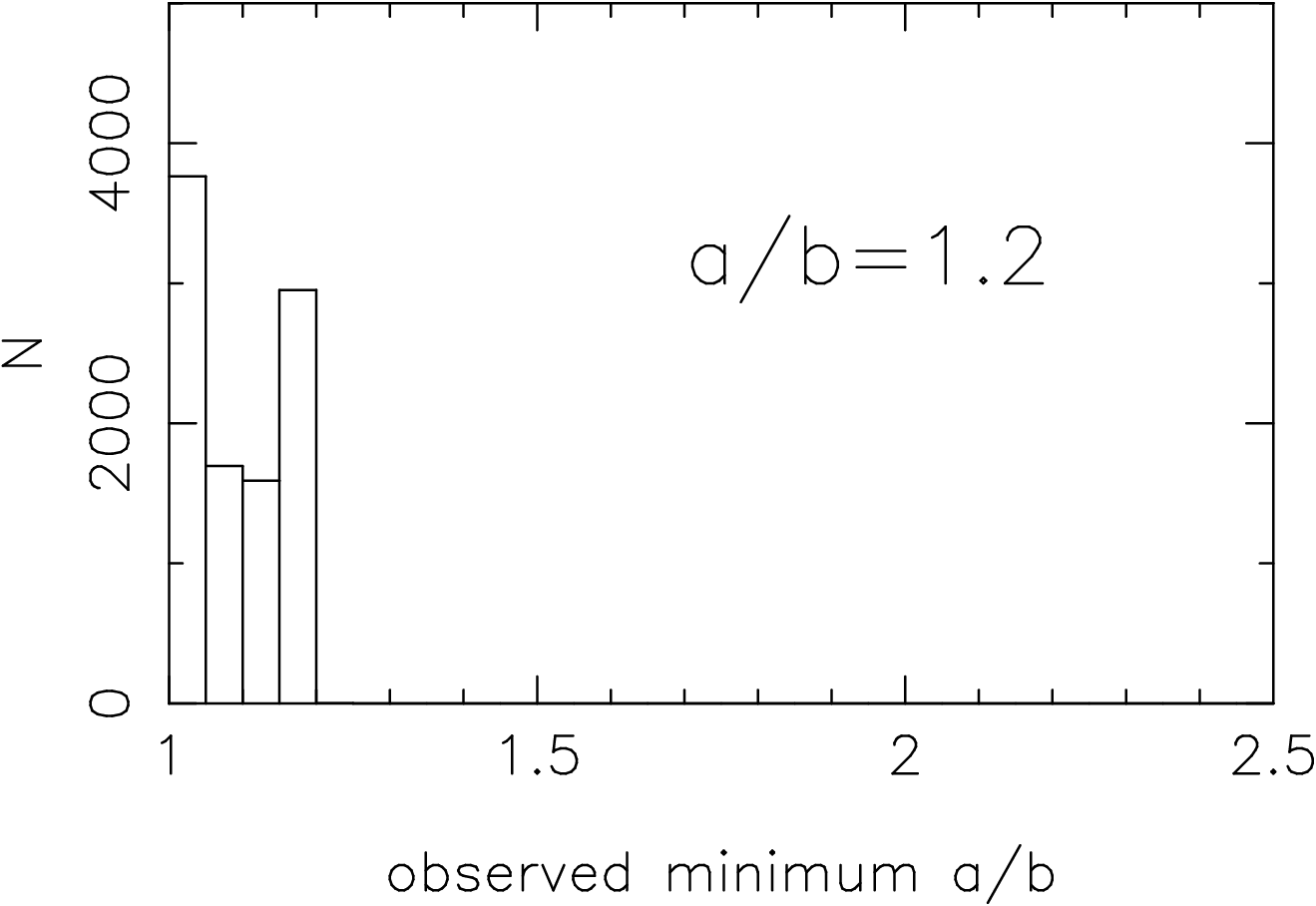} &
   \includegraphics[angle=-90,width=0.22\textwidth]{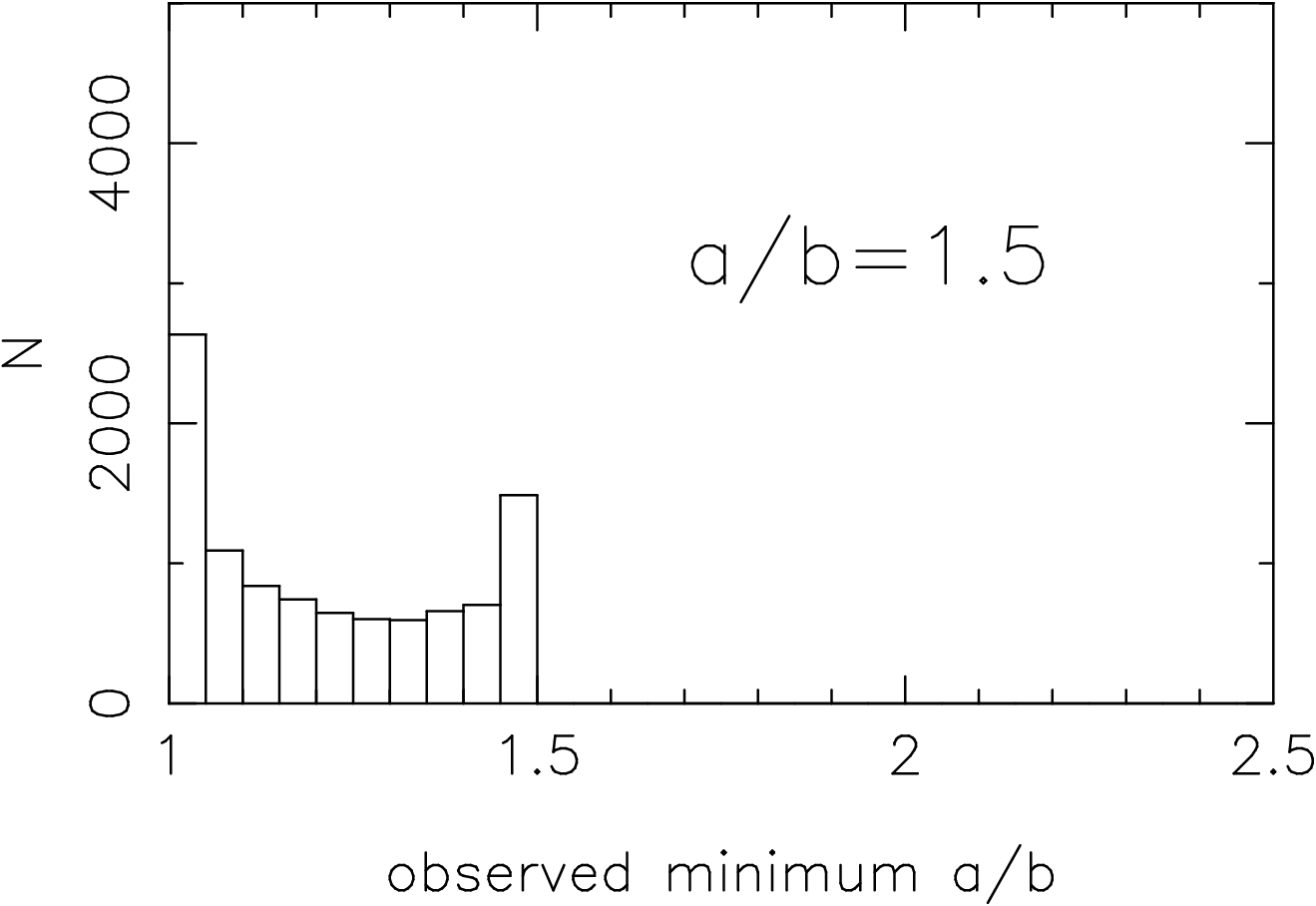}  \\
   \includegraphics[angle=-90,width=0.22\textwidth]{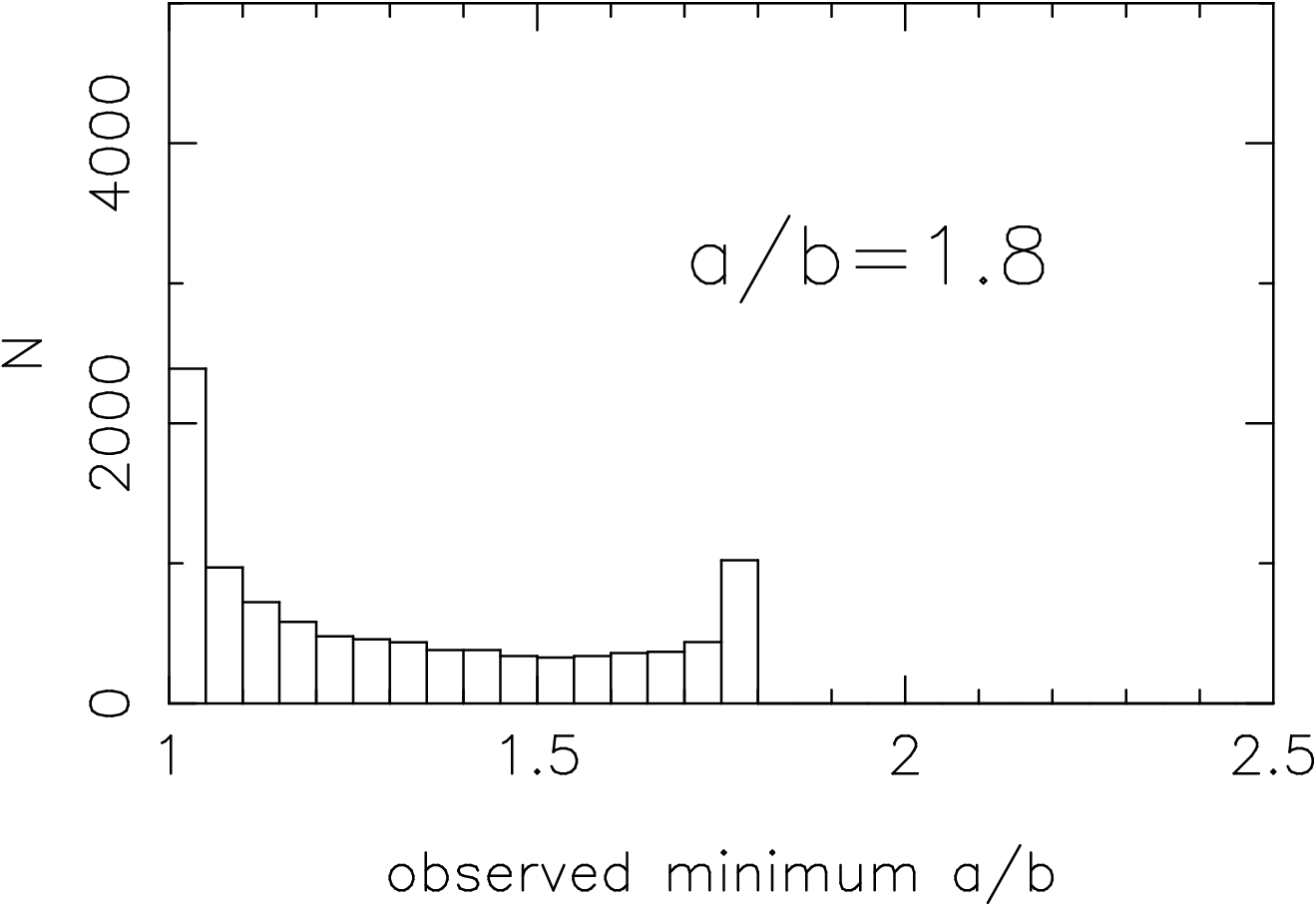} &
   \includegraphics[angle=-90,width=0.22\textwidth]{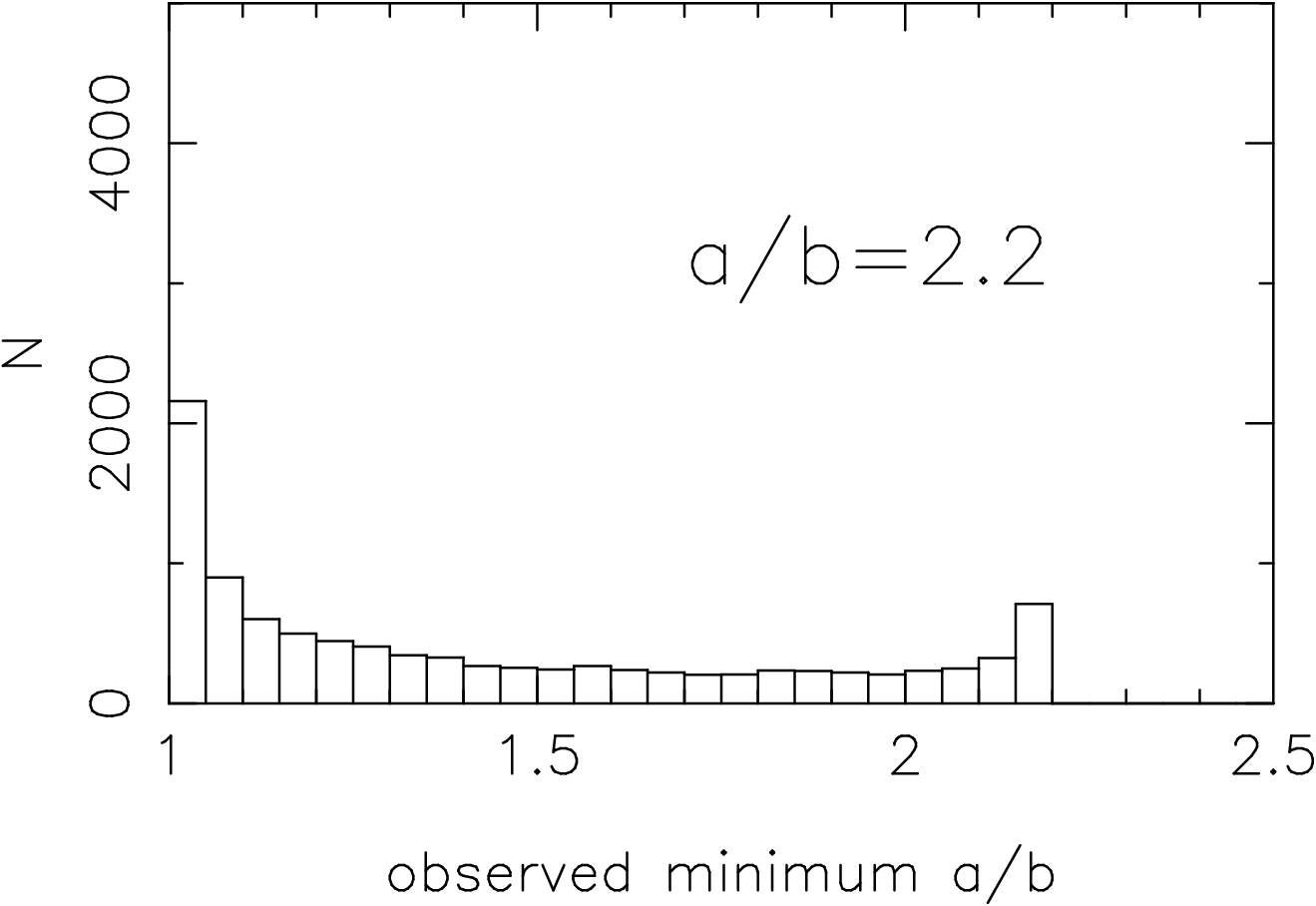}  \\
\end{tabular}
   \caption{Distributions of apparent minimum $a/b$ for 10,000 `observations' of nuclei with fixed axial-ratio at random $\epsilon$. Input values of true $a/b = 1.2$, 1.5, 1.8, 2.2.}
   \label{fig:abdist_fixed}
\end{figure}

In order to describe how the observed minimum $a/b$ is related to the true axial ratio of the nucleus, we substitute equation \ref{area_eqn} into the ratio of areas given in equation \ref{delta_m_eqn}:
\begin{equation}\label{effective_axial_ratio}
\left(\frac{a}{b}\right)_{\rm min} = \frac{S(\phi=0\degr)}{S(\phi=90\degr)} = \left[\left(\frac{a}{b}\right)^{-2}_{\rm true}\sin^2{\epsilon} + \cos^2{\epsilon}\right]^{-\frac{1}{2}}
\end{equation}
This returns the true $a/b$ when the nucleus is viewed from the equator ($\epsilon=90\degr$) and no light curve at all (observed $a/b=1$) from the pole ($\epsilon=0\degr$). The form of this equation means that the observed minimum $a/b$ increases slowly from 1 as $\epsilon$ increases from 0, rising steeply towards the true value at $\epsilon \ga 60\degr$ for highly elongated nuclei (see fig.~1 of \citet{Lamy-chapter}). A histogram of the observed minimum $a/b$ for a sequence of light curves of an object, taken at random values of $\epsilon$, is bimodal. It has a larger peak at $a/b \approx 1$ and a smaller one at the true $a/b$ (fig. \ref{fig:abdist_fixed}). We show histograms for four different nuclei with axial ratios of 1.2, 1.5, 1.8 and 2.2. As the true $a/b$ increases, the difference in the relative size of the peaks in the histogram also increases; for nuclei with elongations $\ga 1.5$ a constant $\sim$ 20\% of the `observed' light curves give a minimum $a/b \approx 1$. The number of observations that return the true axial ratio decreases rapidly as the nucleus becomes more elongated, as the probability of seeing a more elongated body near enough to end on to get an accurate measurement of the true minimum area decreases. 

\begin{figure} 
   \centering
 \begin{tabular}{cc}
   \includegraphics[angle=-90,width=0.22\textwidth]{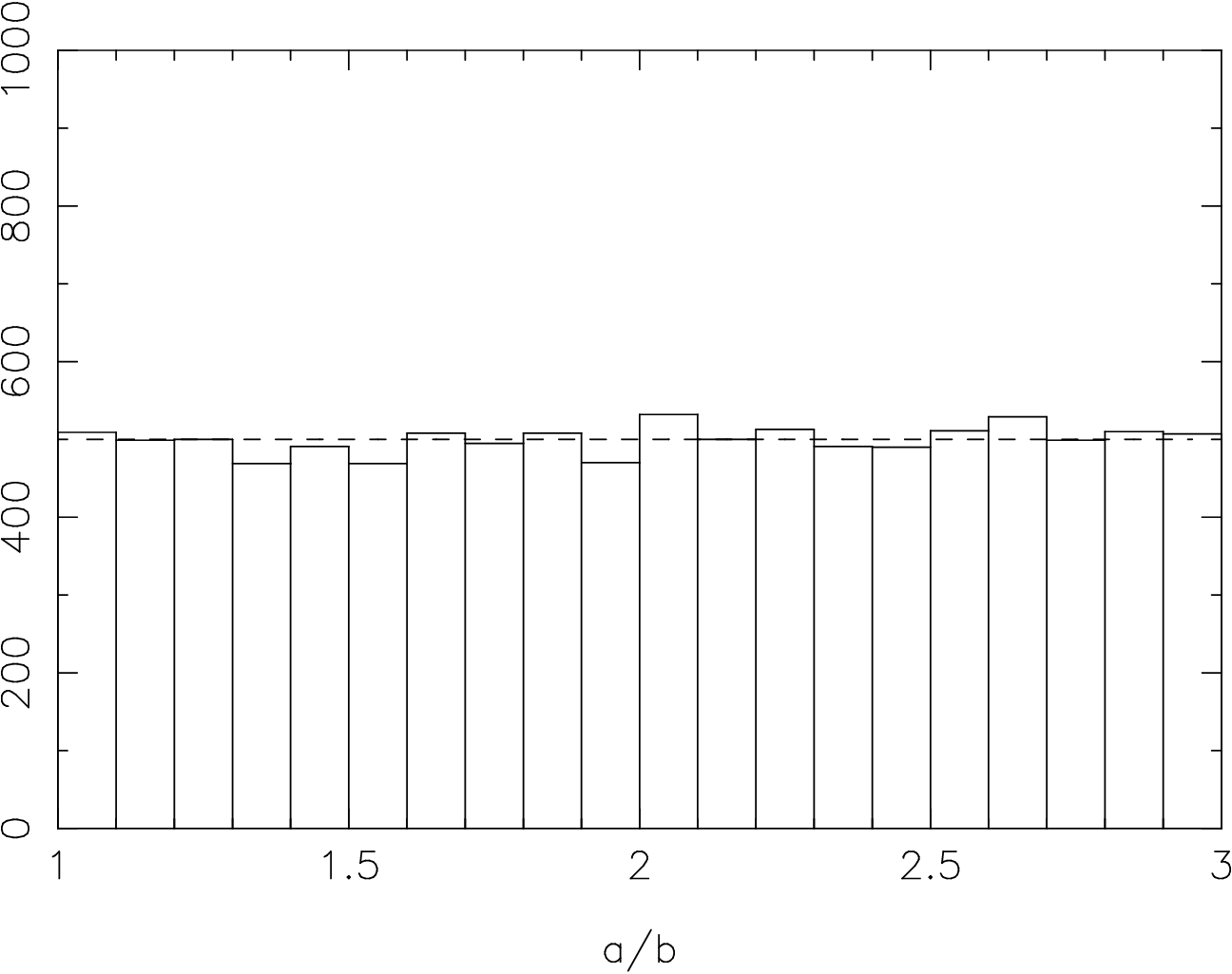} &
   \includegraphics[angle=-90,width=0.22\textwidth]{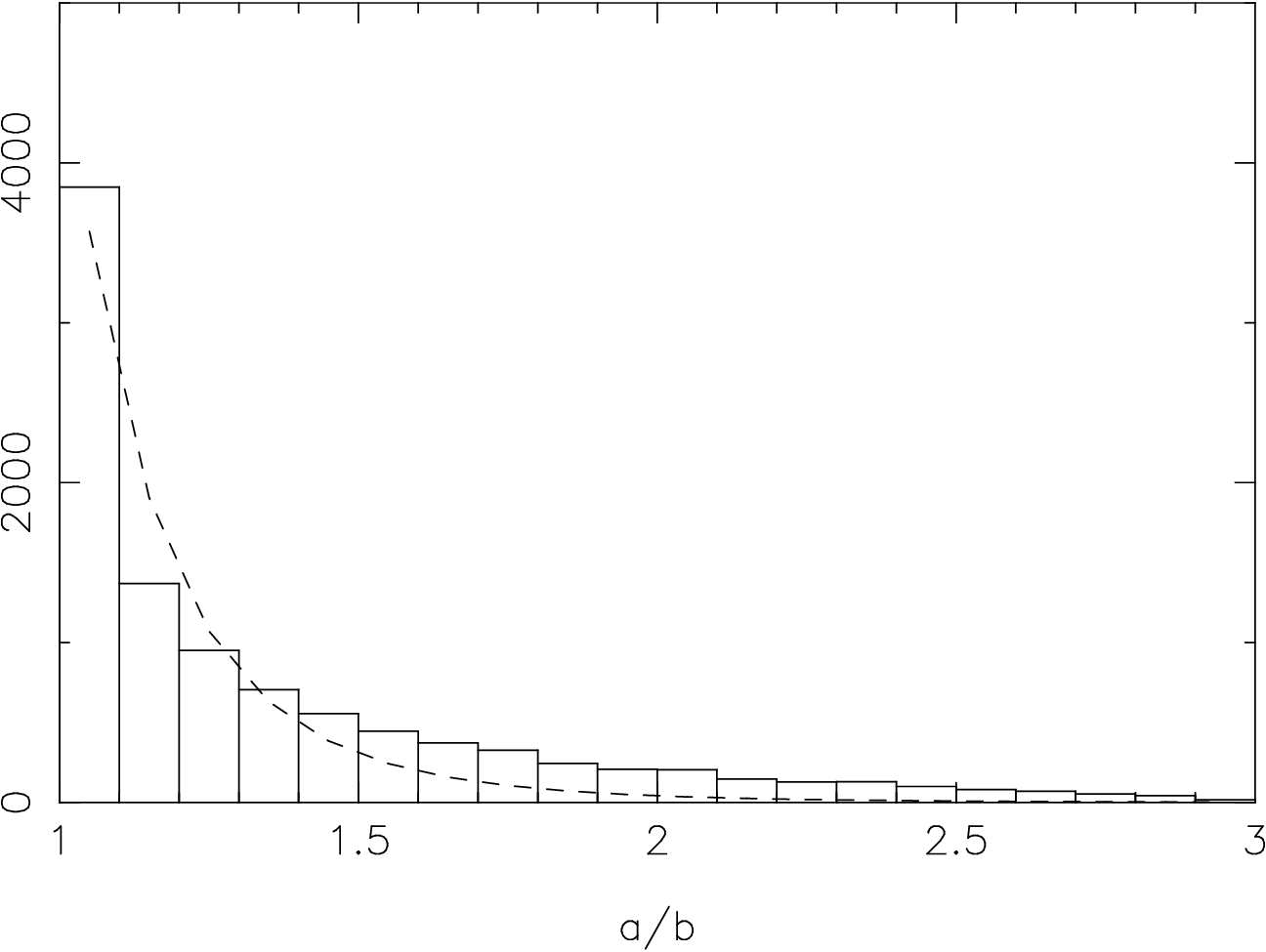}  \\
   \includegraphics[angle=-90,width=0.22\textwidth]{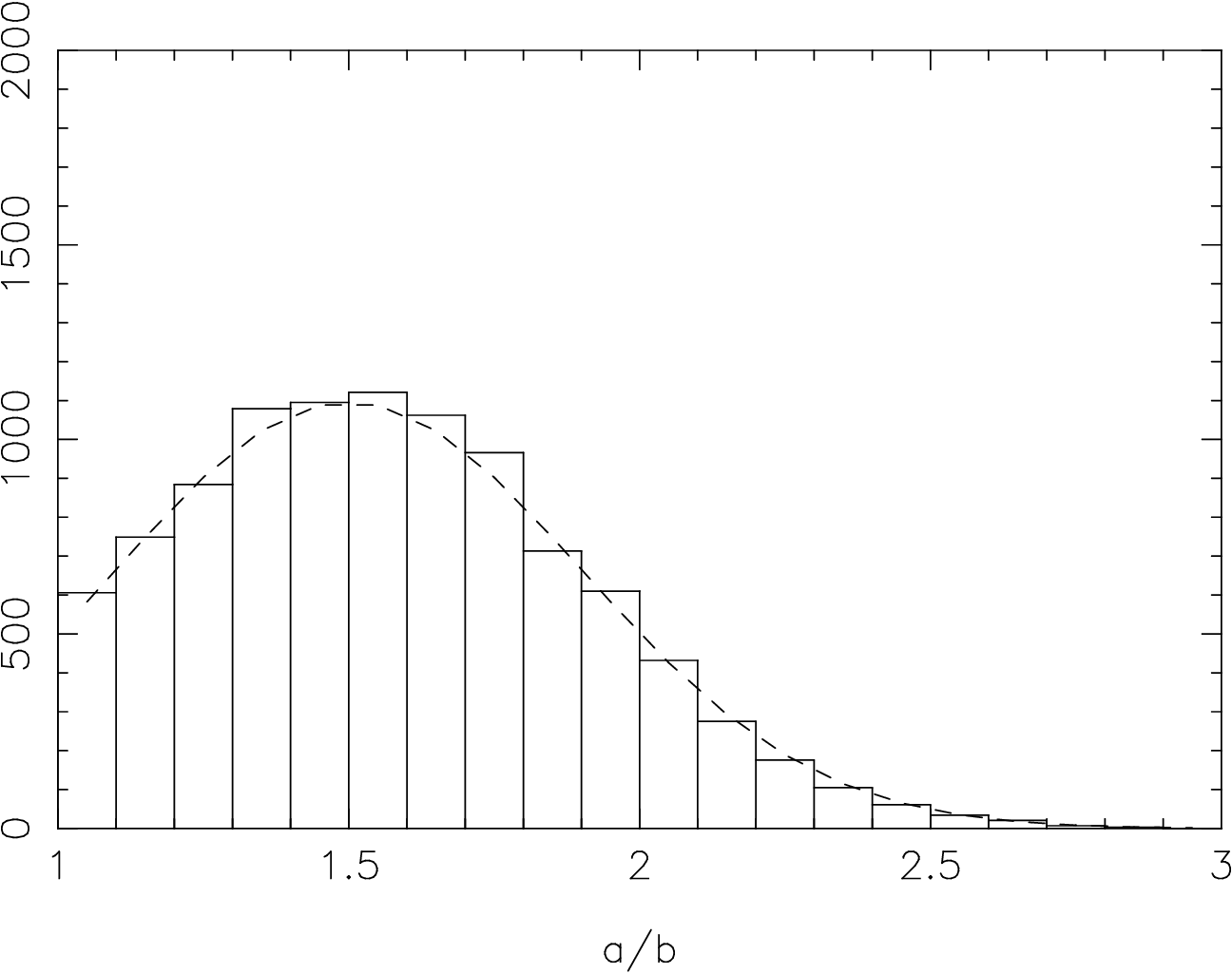} &
   \includegraphics[angle=-90,width=0.22\textwidth]{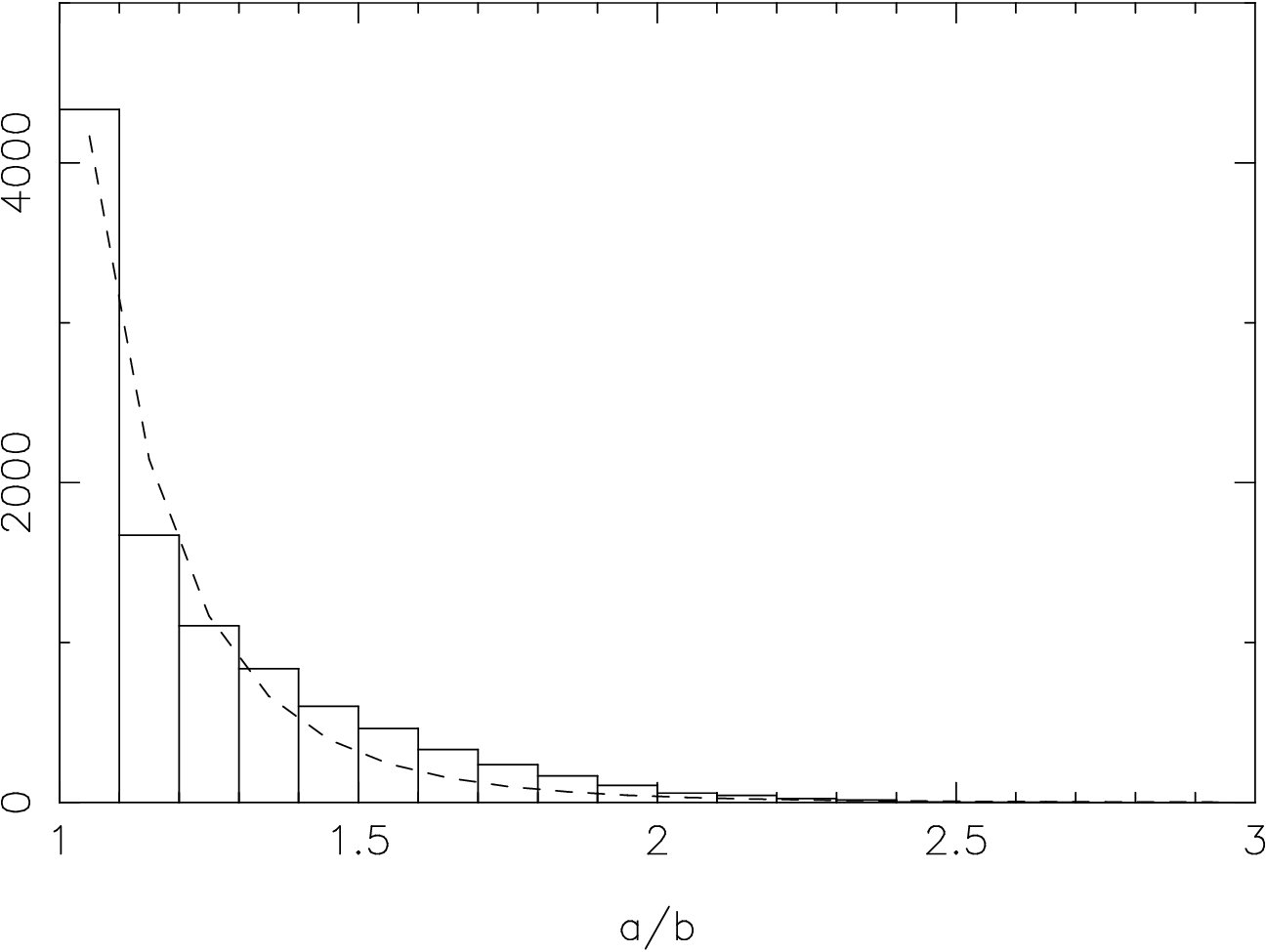}  \\
   \includegraphics[angle=-90,width=0.22\textwidth]{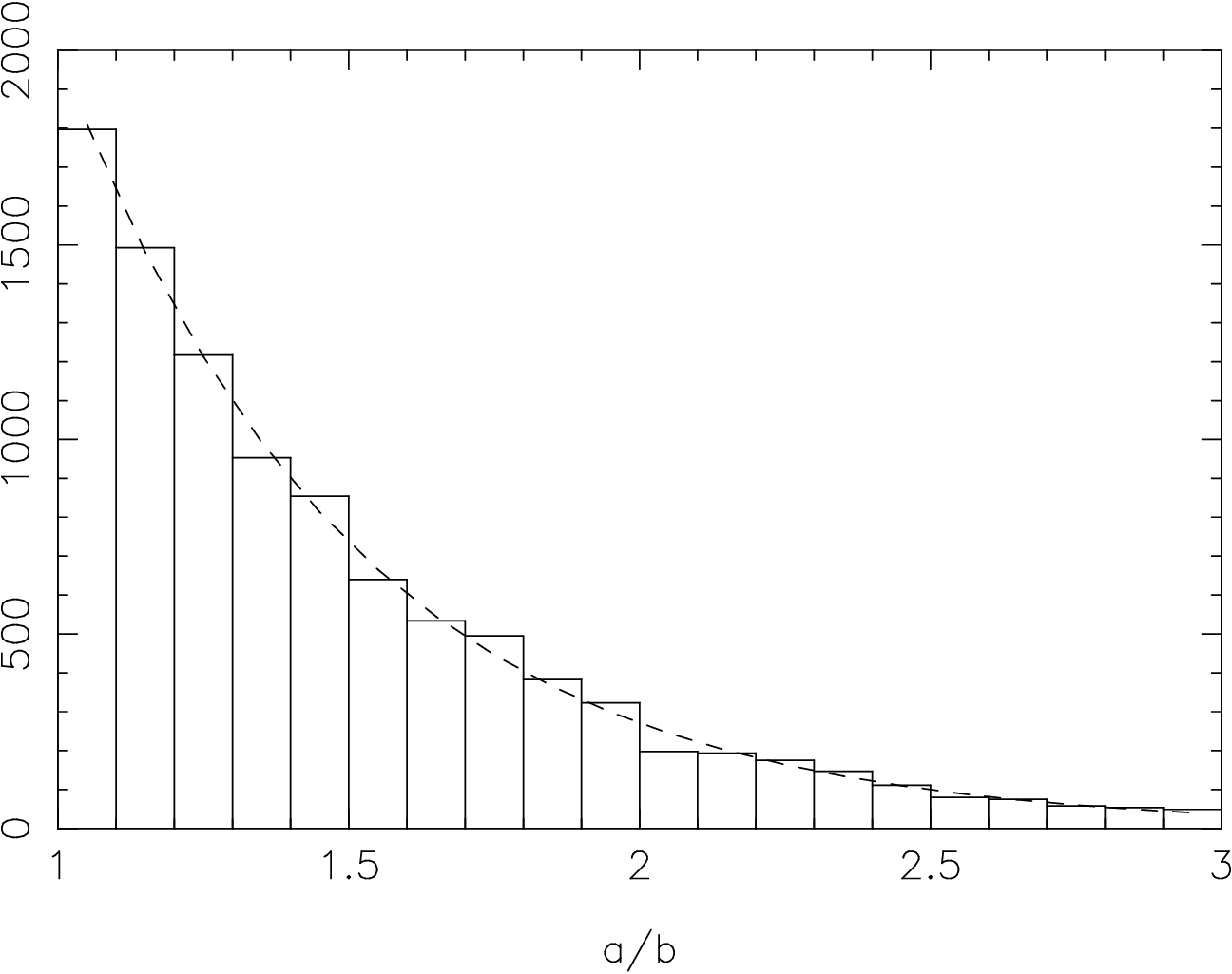} &
   \includegraphics[angle=-90,width=0.22\textwidth]{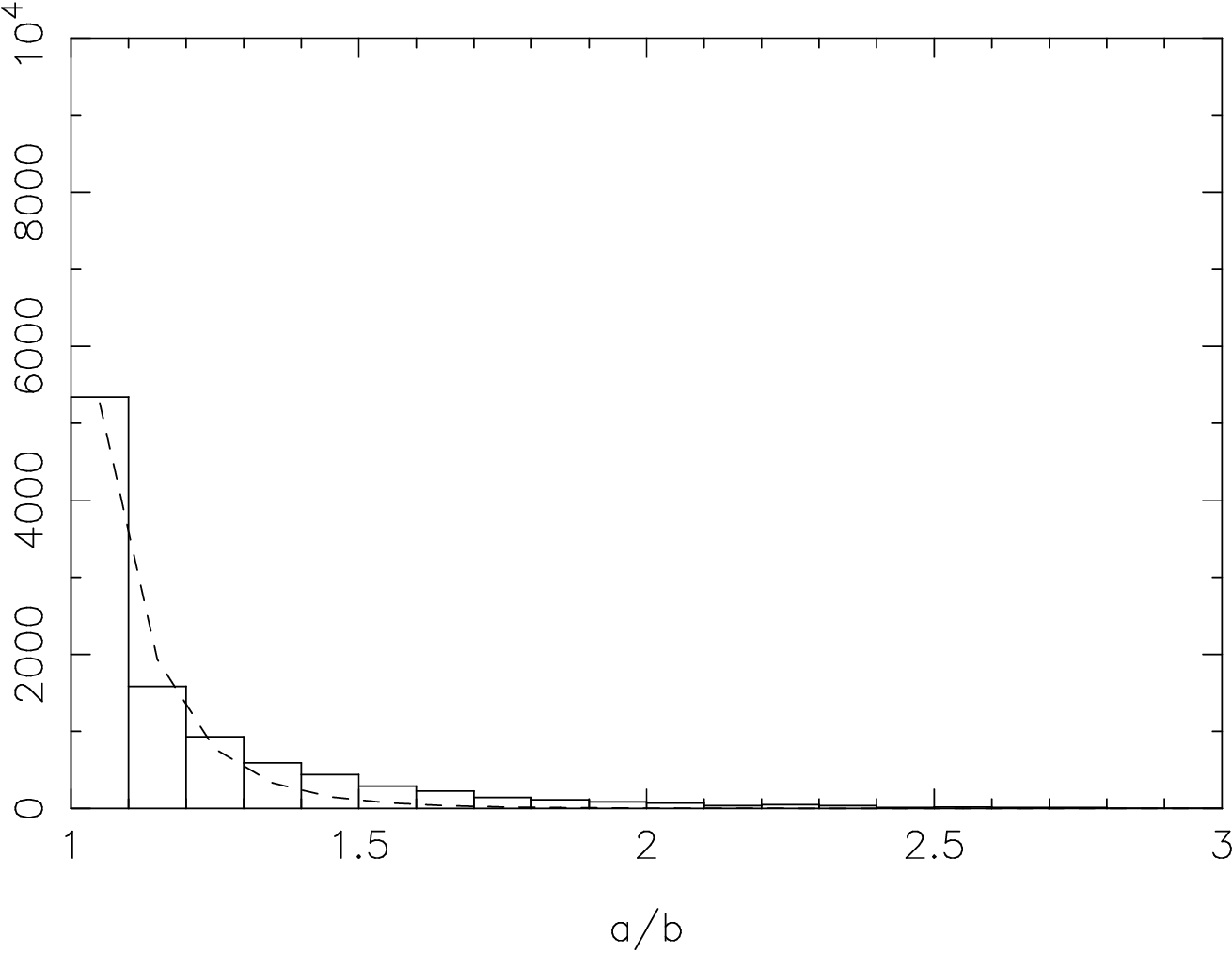}  \\
   \includegraphics[angle=-90,width=0.22\textwidth]{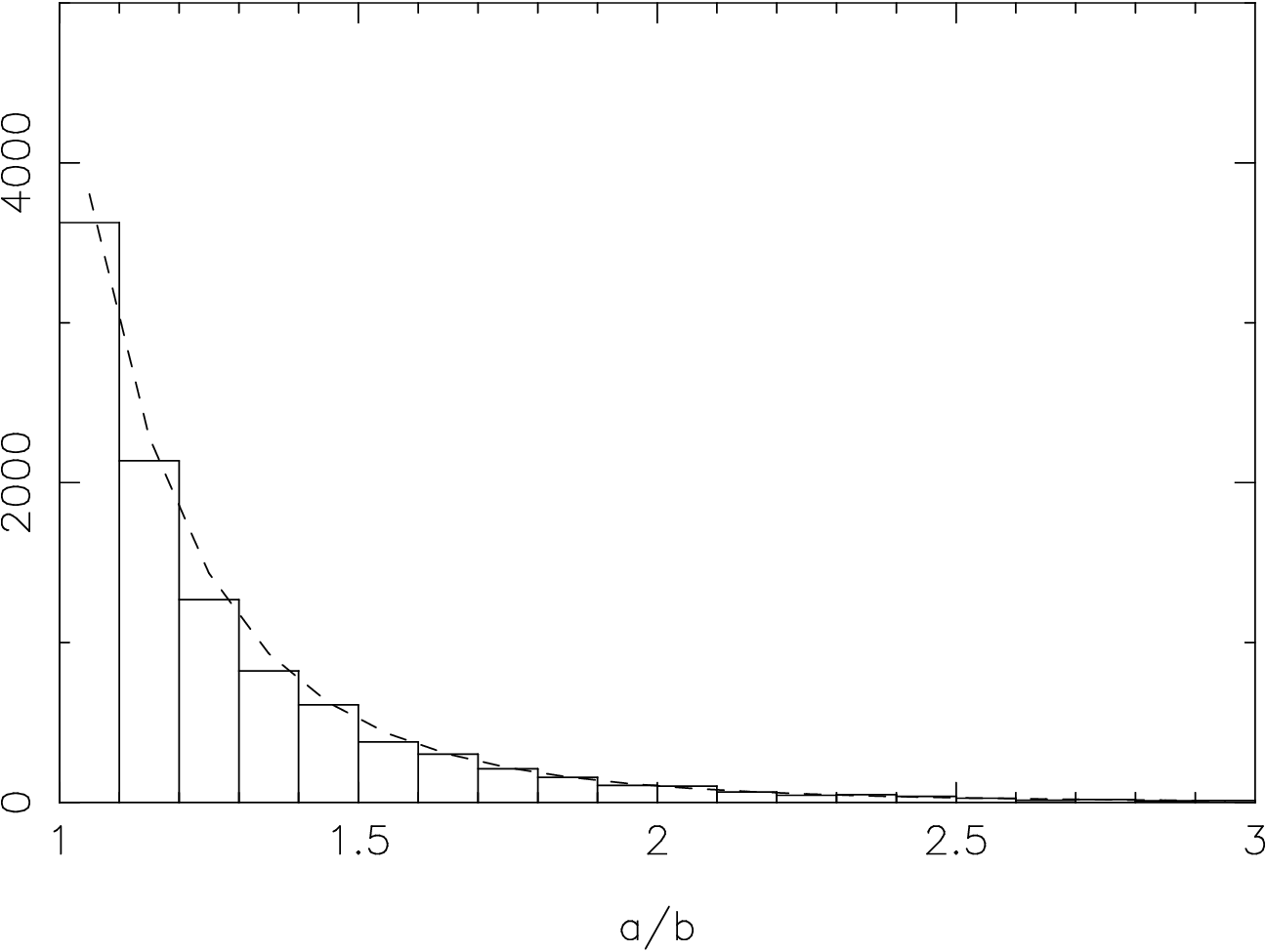} &
   \includegraphics[angle=-90,width=0.22\textwidth]{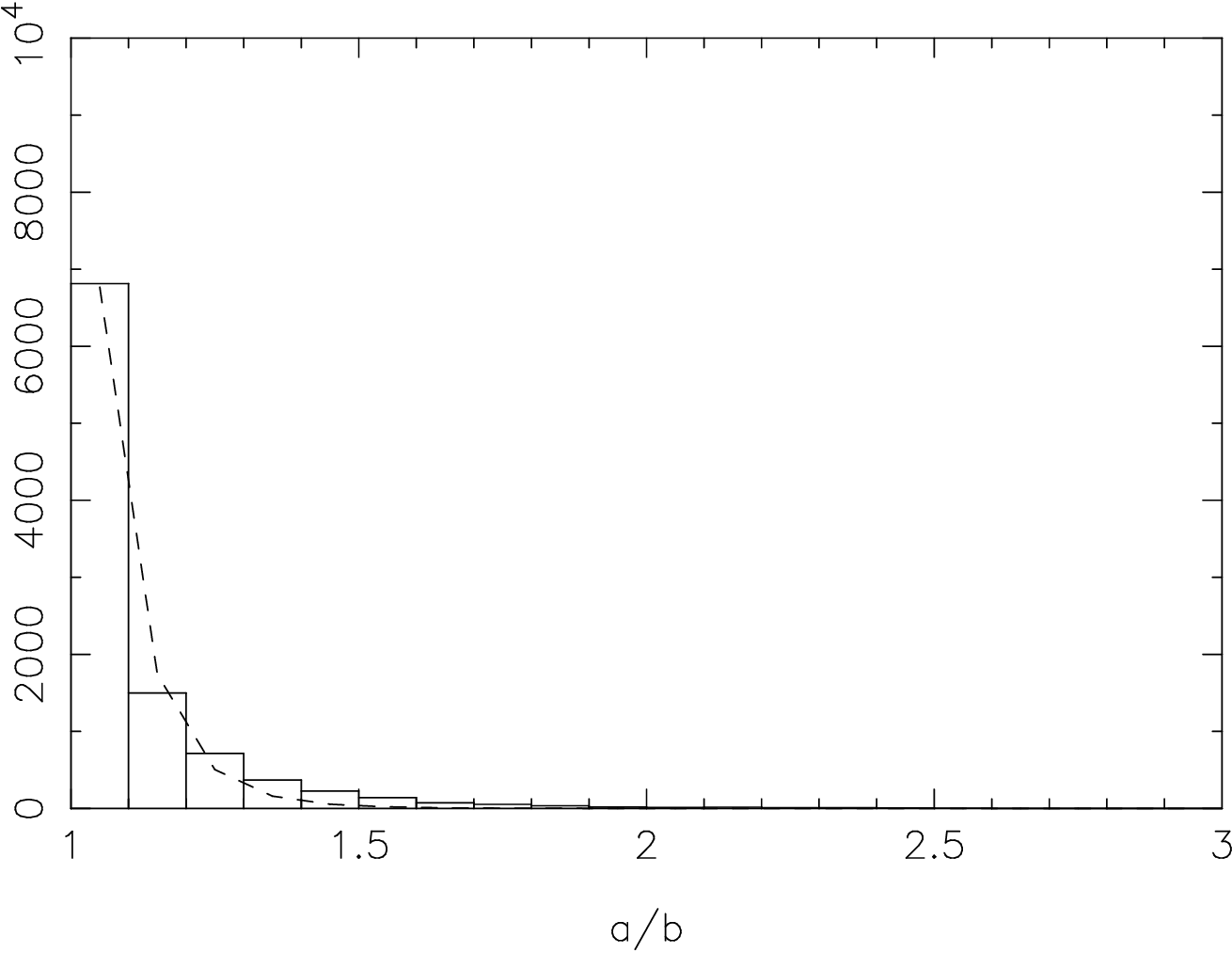}  \\
   \includegraphics[angle=-90,width=0.22\textwidth]{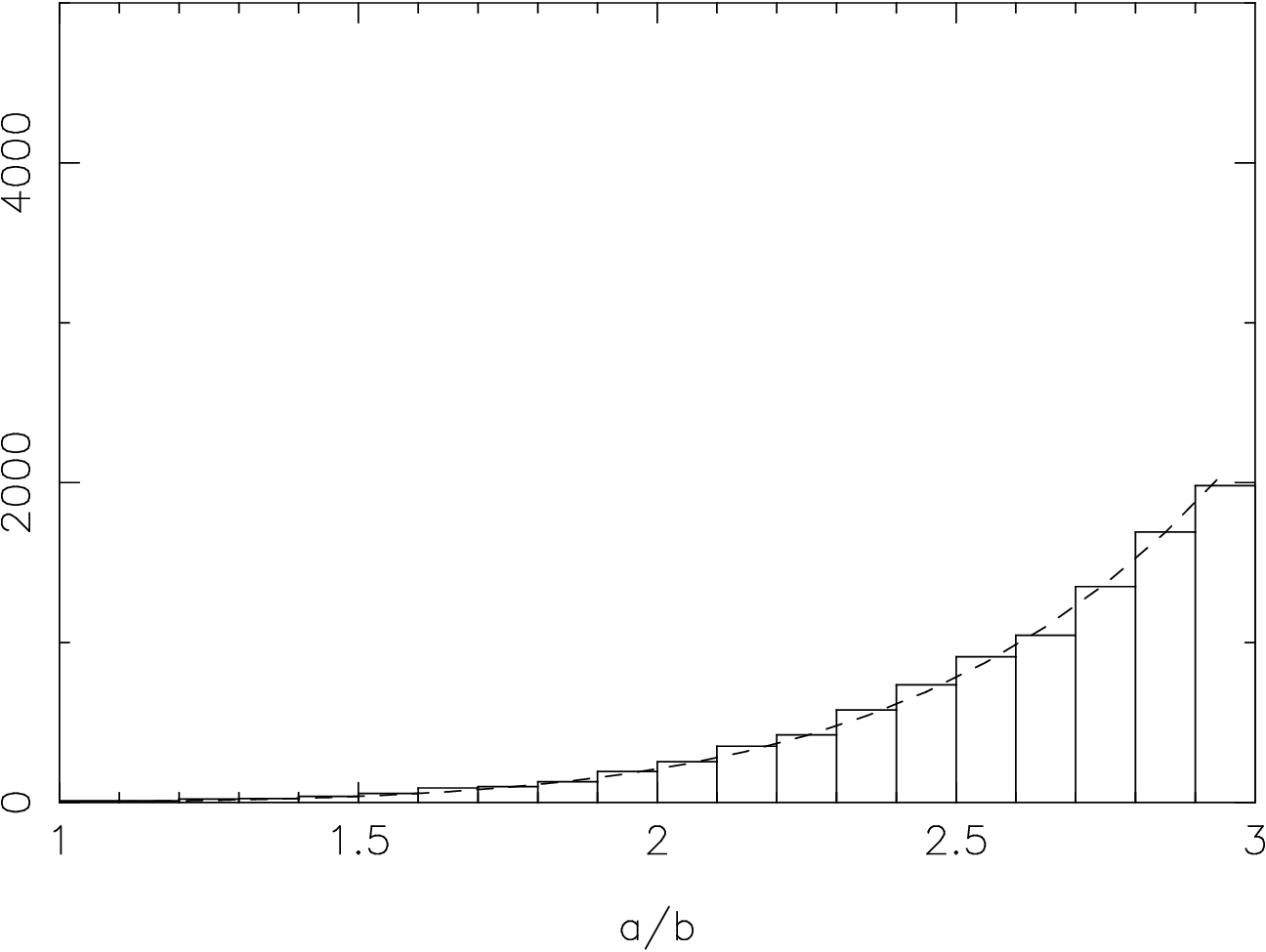} &
   \includegraphics[angle=-90,width=0.22\textwidth]{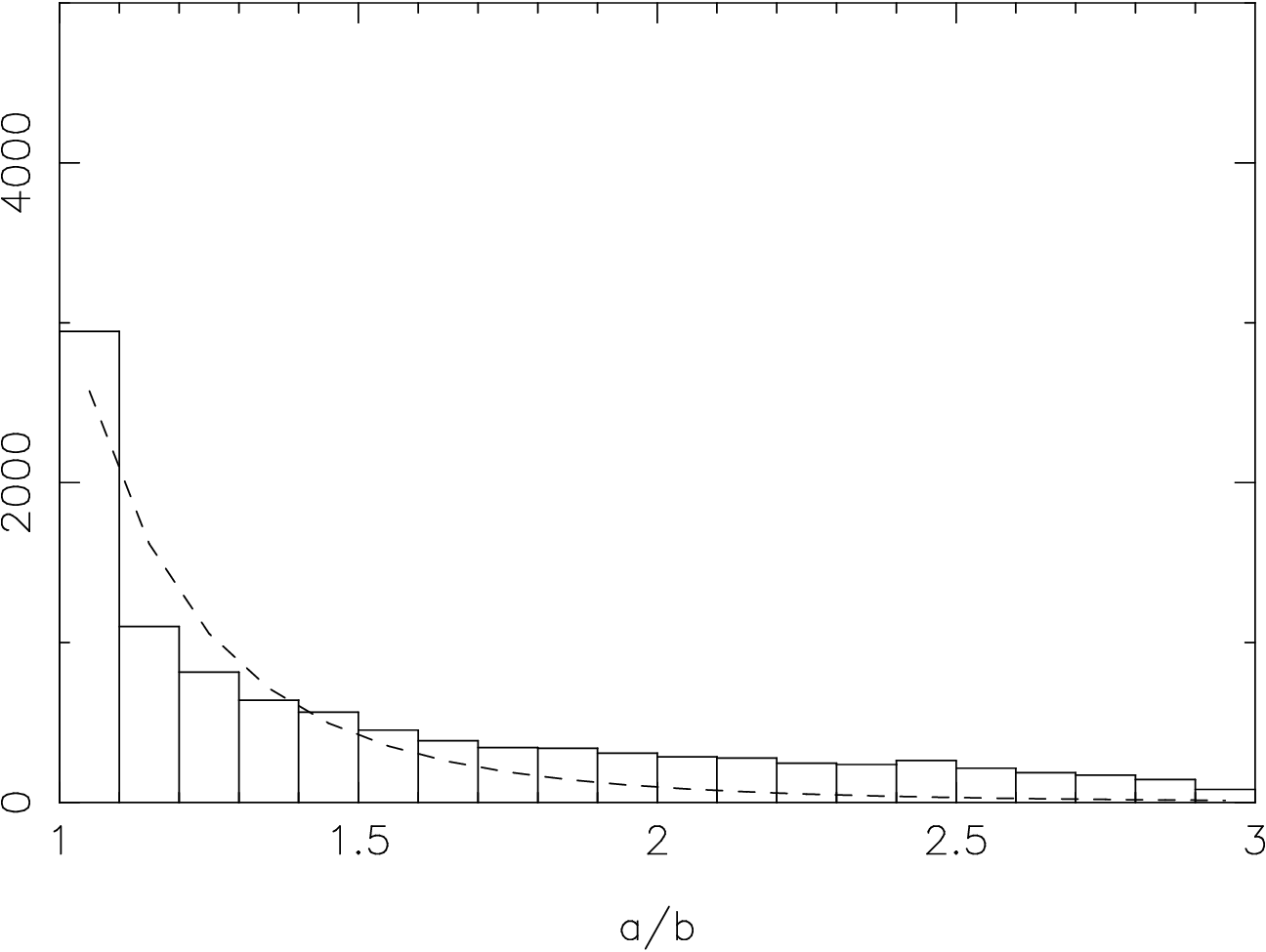}  \\
\end{tabular}
   \caption{Left: Input `true' $a/b$ distributions. Right: Output `observed' distributions based on random $\epsilon$. From the top: Flat distribution between $1 \le a/b \le 3$; truncated Gaussian distribution ($<a/b> = 1.5$; $\sigma=0.6$); exponential distribution ($\tau=0.5$); power law ($x=5.6$); power law ($x=-5.6$). }
   \label{fig:abdist_theory}
\end{figure}

For a collection of nuclei with varying shapes, the distribution of observed minimum $a/b$ from light curves at random $\epsilon$ is a combination of these histograms. We modelled this by generating an input distribution of true shapes and finding the resulting minimum $a/b$ distribution `observed' in $N=10,000$ light curves. We find that for \emph{any} input distribution of true axial-ratios, the randomisation of pole orientation means that light curves return a distribution of observed minimum $a/b$ that is strongly peaked at $\sim 1$ and falls off exponentially. Testing a number of input distributions (flat, Gaussian, exponentially decreasing and power laws with decreasing and even increasing slope) all produced qualitatively similar output distributions, with only slight differences in the tail of the distributions visible in fig. \ref{fig:abdist_theory}. The best fit power laws ($d N/d (a/b) \propto (a/b)^{-x}$) to the output distributions have $x \approx 7$, 7, 11, 15, and 5 respectively. Therefore it is impossible to constrain the underlying shape distribution from the distribution of the minimum axial-ratios observed in light curves. 

In reality, there is an observational bias as nuclei with $a/b_{\rm min} \approx 1$ produce very small amplitude brightness variations, making the light curve un-measureable; it is impossible to know that the full range of the light curve has been observed if peaks and troughs are not recognisable, unless the rotation period of the nucleus is independently known. The difference between the appearance of the theoretical `observed' distributions and that of fig.~\ref{fig:abd_obs} is explained by this lack of reported light curves with $\Delta m \approx 0$. 

\subsubsection{Assumed distribution and MC results}

\begin{figure} 
   \centering
 \begin{tabular}{cc}
   \includegraphics[angle=-90,width=0.22\textwidth]{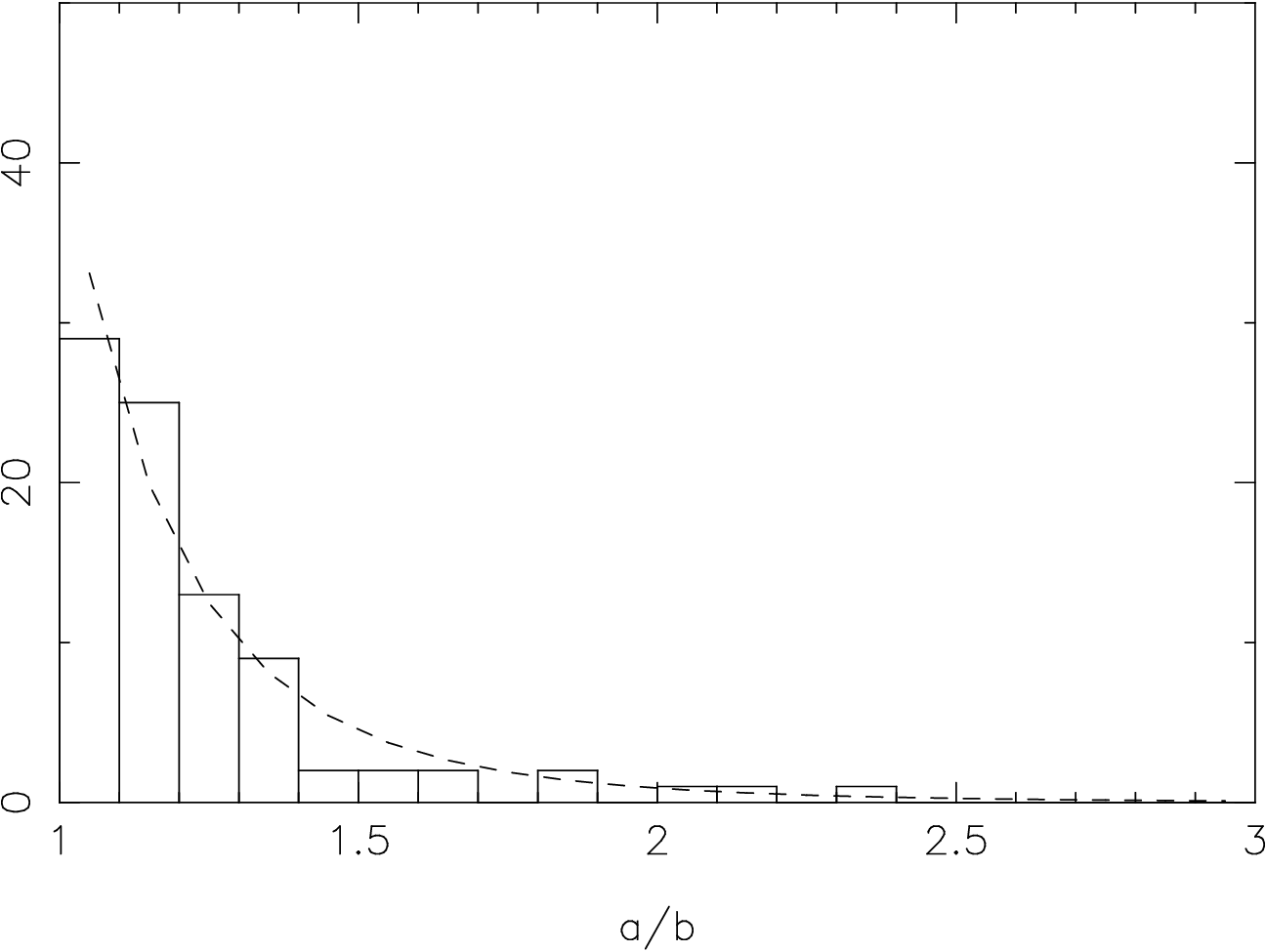} &
   \includegraphics[angle=-90,width=0.22\textwidth]{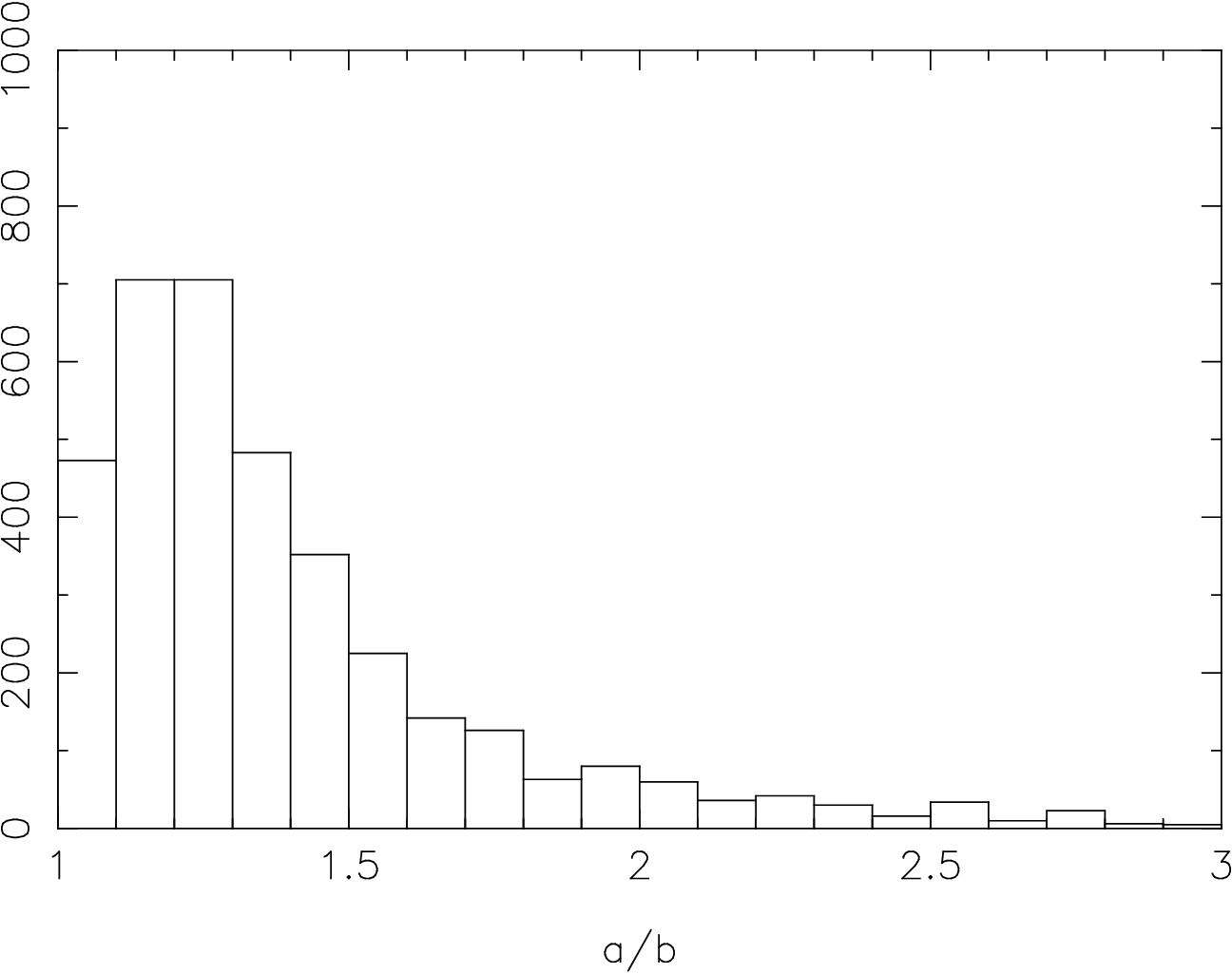} \\
\end{tabular}
   \caption{Shape distribution for asteroids. Left: True $a/b$ distribution from shape models \citep{Torppa08}. Right: Observed $a/b$ distribution from LCDB.}
   \label{fig:abd_asteroids}
\end{figure}

As we cannot constrain the true distribution of axial ratios from the observed distribution, we are forced to assume a distribution for our M-C simulations: We choose to take asteroid shapes as a reasonable basis. The observed distribution of minimum axial ratios from 3666 light curves from the Asteroid Light Curve Data Base \citep[LCDB: ][]{LCDB} shows the same expected shape: A decrease in $N$ as $a/b$ increases but with a lack of objects at $a/b \approx 1$ where the light curve cannot be measured (fig. \ref{fig:abd_asteroids}). However, for asteroids we can compare this with the distribution of `true' shapes as a significant number have detailed shape models. \citet{Torppa08} published a list of simplified descriptions of 87 models; the distribution of axial ratios from this work is best described by a power law with $x=5.6$ (fig. \ref{fig:abd_asteroids}), although we note that this sample is dominated by large asteroids. 

Using this power law ($x=5.6$, normalised over the range $1 \le a/b \le 3$) gives $q= 2.01 \pm 0.07$. Changing the normalisation to $1 \le a/b \le 5$ makes no difference. There is also only a very small difference ($\Delta q<0.05$) for a range of power law distributions of the axial ratio from $x=2.6$ to 8.6, although the uncertainty on the fit increases with decreasing $x$. This is most likely due to the smaller variation in shapes (and hence smaller variation in corrected size) at larger $x$, where a larger proportion of bodies are approximately spherical.

We also tested other shape distributions. Using a Gaussian distribution of $a/b$ (average 1.5, $\sigma=0.6$; a reasonable approximation to the observed distribution of \emph{minimum} $a/b$ values) instead gives $q=1.95\pm 0.10$. Changing the width of the distribution has no appreciable effect on $q$, while increasing the mean to $a/b = 2.0$ slightly decreases the slope to $q=1.91 \pm 0.11$. A flat distribution of $a/b$ values between 1 and 3 gives $q= 1.92 \pm 0.12$, extending it to $a/b=5$ gives $q= 1.87 \pm 0.15$. In general, the slope of the CSD decreases as the distribution includes more elongated bodies, but due to the geometrical effects described above, there is little difference for any reasonable shape distribution. While we have shown that it is impossible to constrain the underlying shape distribution from light curve data alone due to random pole orientations, the same effect means that snap-shots tend to return approximately the correct effective radius without any knowledge of the true shape. Therefore the contribution to the uncertainty on the CSD due to the unknown shape of nuclei is not very large.

\subsection{Including all parameters}

A Monte Carlo run allowing all parameters to vary within our preferred distributions for each (Gaussians with the observed mean and $\sigma$ values for $A_R$ and $\beta$, and an $x=5.6$ power law $a/b$ distribution) represents our best description of what we expect the true size distribution to be, and gives an uncertainty including all parameters. The fit to this has $q = 1.92 \pm 0.20$, with an average cut off in radius at 1.25 km, and a best fit shallow slope of $0.18 \pm 0.03$ at sizes below this. The same run but fixing $\beta$ and $A_R$ at the `known' values for the subset of comets with measurements gives $q=2.01\pm0.20$. Note the relatively large uncertainty when including all parameters at once; we take this as a fair assessment of the true uncertainty on the CSD slope including all unknown factors.

\section{Discussion}

\begin{table*}
\begin{minipage}{156mm}
\caption{Inputs and results from Monte Carlo runs.}
\begin{center}
\begin{tabular}{l|c l l p{2.7cm} c c|c c c}
\hline
\# & Mag\footnote{%
Vary input magnitude within photometric error bars?
} & Phase function & Albedo & Shape\footnote{%
Assumed shape distribution for nuclei.
} & Fix\footnote{%
Use fixed values for $\beta$ and $A_R$ when known for a given comet? If no, all comets have values selected from the specified distribution, even those where these values are independently constrained.
} & $N$ & $q_1$\footnote{%
$q_1$ is the best fit at $r_{\rm N} \le$ cut-off.
}  & $q_2$\footnote{%
$q_2$ is the best fit at $r_{\rm N} >$ cut-off, i.e. the quoted $q$.
} & Cut-off\\
\hline
\multicolumn{4}{l}{\it Fixed assumed values:}\\
1 & N & 0.035 & 0.04 & Spherical & N & 1 & 0.19 & 1.97 & 1.26 \\ 
2 & N & 0.035 & 0.04 & Spherical & Y & 1 & 0.19 & 2.00 & 1.26 \\ 
\multicolumn{4}{l}{\it Adjust photometry:}\\
3 & Y & 0.035 & 0.04 & Spherical & N & 1000 & 0.18 $\pm$0.01 & 1.93 $\pm$0.10 & 1.24 $\pm$0.07 \\ 
\multicolumn{4}{l}{\it Adjust phase function:}\\
4 & N & 0.025 & 0.04 & Spherical & N & 1 & 0.18 & 1.94 & 1.17 \\ 
5 & N & 0.045 & 0.04 & Spherical & N & 1 & 0.19 & 1.98 & 1.35 \\ 
6 & N & 0.065 & 0.04 & Spherical & N & 1 & 0.20 & 1.99 & 1.50 \\ 
7 & N & 0.053 $\pm$ 0.016 & 0.04 & Spherical & N & 1000 & 0.19 $\pm$0.01 & 1.95 $\pm$0.07 & 1.38 $\pm$0.07 \\ 
8 & N & 0.053 $\pm$ 0.005 & 0.04 & Spherical & N & 1000 & 0.20 $\pm$0.00 & 1.99 $\pm$0.02 & 1.42 $\pm$0.03 \\ 
9 & N & 0.053 $\pm$ 0.025 & 0.04 & Spherical & N & 1000 & 0.19 $\pm$0.02 & 1.91 $\pm$0.11 & 1.35 $\pm$0.09 \\ 
10 & N & 0.053 $\pm$ 0.016 & 0.04 & Spherical & Y & 1000 & 0.19 $\pm$0.01 & 2.05 $\pm$0.06 & 1.39 $\pm$0.06 \\ 
11 & N & 0.010 $\pm$ 0.010 & 0.04 & Spherical & N & 1000 & 0.15 $\pm$0.01 & 1.72 $\pm$0.06 & 1.02 $\pm$0.06 \\ 
12 & N & 0.020 $\pm$ 0.010 & 0.04 & Spherical & N & 1000 & 0.16 $\pm$0.02 & 1.81 $\pm$0.11 & 1.13 $\pm$0.11 \\ 
13 & N & 0.030 $\pm$ 0.010 & 0.04 & Spherical & N & 1000 & 0.18 $\pm$0.01 & 1.94 $\pm$0.09 & 1.26 $\pm$0.09 \\ 
14 & N & 0.040 $\pm$ 0.010 & 0.04 & Spherical & N & 1000 & 0.19 $\pm$0.01 & 1.97 $\pm$0.05 & 1.32 $\pm$0.06 \\ 
15 & N & 0.050 $\pm$ 0.010 & 0.04 & Spherical & N & 1000 & 0.19 $\pm$0.01 & 1.97 $\pm$0.04 & 1.38 $\pm$0.06 \\ 
16 & N & 0.060 $\pm$ 0.010 & 0.04 & Spherical & N & 1000 & 0.20 $\pm$0.01 & 1.97 $\pm$0.05 & 1.43 $\pm$0.06 \\ 
17 & N & 0.070 $\pm$ 0.010 & 0.04 & Spherical & N & 1000 & 0.20 $\pm$0.01 & 1.95 $\pm$0.05 & 1.47 $\pm$0.06 \\ 
18 & N & 0.080 $\pm$ 0.010 & 0.04 & Spherical & N & 1000 & 0.20 $\pm$0.01 & 1.92 $\pm$0.05 & 1.49 $\pm$0.05 \\ 
19 & N & 0.090 $\pm$ 0.010 & 0.04 & Spherical & N & 1000 & 0.20 $\pm$0.01 & 1.90 $\pm$0.04 & 1.54 $\pm$0.04 \\ 
\multicolumn{4}{l}{\it Adjust albedo:}\\
20 & N & 0.035 & 0.044 $\pm$ 0.013 & Spherical & N & 1000 & 0.17 $\pm$0.02 & 1.85 $\pm$0.15 & 1.16 $\pm$0.10 \\ 
21 & N & 0.035 & 0.020 $\pm$ 0.013 & Spherical & N & 1000 & 0.16 $\pm$0.03 & 1.64 $\pm$0.21 & 1.64 $\pm$0.21 \\ 
22 & N & 0.035 & 0.060 $\pm$ 0.013 & Spherical & N & 1000 & 0.18 $\pm$0.02 & 1.89 $\pm$0.12 & 1.01 $\pm$0.08 \\ 
23 & N & 0.035 & 0.040 $\pm$ 0.005 & Spherical & N & 1000 & 0.18 $\pm$0.01 & 1.93 $\pm$0.09 & 1.26 $\pm$0.08 \\ 
24 & N & 0.035 & 0.040 $\pm$ 0.020 & Spherical & N & 1000 & 0.16 $\pm$0.03 & 1.72 $\pm$0.19 & 1.18 $\pm$0.13 \\ 
25 & N & 0.035 & 0.044 $\pm$ 0.013 & Spherical & Y & 1000 & 0.17 $\pm$0.02 & 1.85 $\pm$0.14 & 1.16 $\pm$0.10 \\ 
\multicolumn{4}{l}{\it Adjust shape:}\\
26 & N & 0.035 & 0.04 & Flat, 1 $\le a/b \le$ 3 & N & 1000 & 0.18 $\pm$0.02 & 1.92 $\pm$0.12 & 1.17 $\pm$0.09 \\ 
27 & N & 0.035 & 0.04 & Flat, 1 $\le a/b \le$ 5 & N & 1000 & 0.17 $\pm$0.02 & 1.87 $\pm$0.15 & 1.09 $\pm$0.09 \\ 
28 & N & 0.035 & 0.04 & Gaussian, 1.5 $\pm$ 0.6 & N & 1000 & 0.18 $\pm$0.01 & 1.95 $\pm$0.10 & 1.22 $\pm$0.08 \\ 
29 & N & 0.035 & 0.04 & Gaussian, 1.5 $\pm$ 0.3 & N & 1000 & 0.18 $\pm$0.01 & 1.96 $\pm$0.08 & 1.23 $\pm$0.07 \\ 
30 & N & 0.035 & 0.04 & Gaussian, 1.5 $\pm$ 0.9 & N & 1000 & 0.18 $\pm$0.01 & 1.94 $\pm$0.11 & 1.20 $\pm$0.08 \\ 
31 & N & 0.035 & 0.04 & Gaussian, 2 $\pm$ 0.6 & N & 1000 & 0.18 $\pm$0.01 & 1.91 $\pm$0.11 & 1.18 $\pm$0.08 \\ 
32 & N & 0.035 & 0.04 & Power law, $x =$ -5.6, 1 $\le a/b \le$ 3 & N & 1000 & 0.18 $\pm$0.01 & 2.01 $\pm$0.07 & 1.26 $\pm$0.06 \\ 
33 & N & 0.035 & 0.04 & Power law, $x =$ -5.6, 1 $\le a/b \le$ 5 & N & 1000 & 0.18 $\pm$0.01 & 2.02 $\pm$0.07 & 1.26 $\pm$0.07 \\ 
34 & N & 0.035 & 0.04 & Power law, $x =$ -4.6, 1 $\le a/b \le$ 5 & N & 1000 & 0.18 $\pm$0.01 & 2.00 $\pm$0.08 & 1.25 $\pm$0.07 \\ 
35 & N & 0.035 & 0.04 & Power law, $x =$ -6.6, 1 $\le a/b \le$ 5 & N & 1000 & 0.18 $\pm$0.01 & 2.02 $\pm$0.06 & 1.27 $\pm$0.06 \\ 
36 & N & 0.035 & 0.04 & Power law, $x =$ -2.6, 1 $\le a/b \le$ 5 & N & 1000 & 0.18 $\pm$0.01 & 1.97 $\pm$0.11 & 1.22 $\pm$0.08 \\ 
37 & N & 0.035 & 0.04 & Power law, $x =$ -8.6, 1 $\le a/b \le$ 5 & N & 1000 & 0.18 $\pm$0.00 & 2.02 $\pm$0.04 & 1.27 $\pm$0.05 \\ 
\multicolumn{4}{l}{\it Adjust all parameters:}\\
38 & Y & 0.035 $\pm$ 0.020 & 0.040 $\pm$ 0.040 & Gaussian, 1.5 $\pm$ 0.6 & N & 10000 & 0.16 $\pm$0.04 & 1.51 $\pm$0.22 & 1.17 $\pm$0.19 \\ 
39 & Y & 0.053 $\pm$ 0.016 & 0.044 $\pm$ 0.013 & Power law, $x =$ -5.6, 1 $\le a/b \le$ 3 & N & 10000 & 0.18 $\pm$0.03 & 1.92 $\pm$0.20 & 1.25 $\pm$0.13 \\ 
40 & Y & 0.053 $\pm$ 0.016 & 0.044 $\pm$ 0.013 & Power law, $x =$ -5.6, 1 $\le a/b \le$ 3 & Y & 10000 & 0.19 $\pm$0.02 & 2.01 $\pm$0.20 & 1.28 $\pm$0.12 \\ 

\hline
\end{tabular}
\end{center}
\end{minipage}
\label{big_MCrun_table}
\end{table*}%

\begin{figure}
   \centering
   \includegraphics[angle=-90,width=0.45\textwidth]{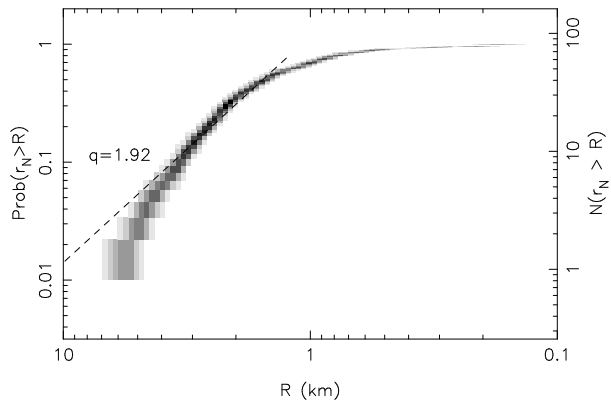} 
   \caption{Probability map for MC run 39, which gives our final result by allowing all parameters to vary. The shading shows the average shape of the 10,000 CSDs: darker areas are the bins in $R$--$P(r_{\rm N} > R)$ space where the majority of CSDs passed through, lighter areas show the outlying areas explored at the ends of the distributions. A dashed line showing the average slope from this run is over-plotted.}
   \label{fig:density}
\end{figure}

Table \ref{big_MCrun_table} shows input parameters and results for all the MC runs, including all of the ones varying only a single parameter at a time described in the previous section and also runs allowing all parameters to vary. This shows that, relative to our reference CSD with $q=1.97$, the largest change  in slope is found when allowing the albedo to vary. This is mostly due to the very shallow slope that results from any albedo distribution that includes exceptionally dark ($A_R \approx 0$) nuclei, which are unlikely to be realistic. In any case, this demonstrates the importance of better constraining the albedo distribution of comets, an important result that the SEPPCoN survey will provide \citep{FernandezY08,Lowry10}. The only other MC runs with a large difference from the reference ($\Delta q \ge 0.1$) are the extreme shape distributions, which are also unlikely to represent reality. We can conclude from this that the uncertainty on the input parameter distributions actually has only a small effect, as the variation in slopes always falls within the typical uncertainty found when varying a single parameter at a time. 

Allowing all parameters to vary within the best-fit distributions gives a final value of $q = 1.92 \pm 0.20$. This value is remarkably close to our reference value, again implying that the combined effect of the various uncertain parameters is actually relatively small. This helps to validate the assumptions made in previous CSD estimates, but the real significance of this work is that the uncertainty on this value has been rigourously determined by including all individual uncertainties on each radius. Figure \ref{fig:density} shows how the CSDs varied in this MC simulation.

Comparing this with the previous CSD estimates listed in Table~\ref{previous} shows that the value we find for $q$ is in a similar range to recent results, if slightly steeper than the consensus value, and when we consider the true uncertainty we find the majority of results are encompassed within $2\sigma$. While previous works assumed constant values for $A_R$ and $\beta$ and did not consider the uncertainties on these, we show that the CSD slope including all sources of uncertainty agrees with these earlier results. It is worth noting that our testing of various distributions for the assumed parameters never produced a slope as steep as the considerably larger $q$ found by \citet{FernandezJ99} and \citet{Tancredi06}, even for extreme distributions we regard as unlikely to be realistic. We suspect that the difference is due to the higher accuracy of the photometric measurements used in our catalogue, all of which come from large aperture telescopes. It is possible that a similar MC analysis of the uncertainty on $q$ due to the photometric uncertainties in the \citet{Tancredi06} catalogue would give a large error bar, making it consistent with the other results, however this catalogue does not include all the necessary information (in its currently published form) to apply our technique.

The CSD slope we find is very close to the theoretical value of $q=2.04$ found by \citet{OBrien+Greenberg03} for a collisionally relaxed population of strengthless bodies. However, the large uncertainty we find means that we cannot claim that this is evidence that JFC nuclei came from collisional disruption of larger bodies; we can simply state that the CSD slope is consistent with that interpretation. It is important to remember that the activity of comets alters their sizes as they lose significant amounts of material with each perihelion passage, and therefore we would not necessarily expect to see a CSD that matched a collisional one, even if the comets are collisional fragments. Both `normal' cometary activity and major mass loss events (nucleus splitting or outbursts) are highly variable between comets and from one orbit to another, so the change in size due to mass loss from a nucleus is a highly complicated process. More detailed theoretical models describing the expected CSD of nuclei for various formation (and evolution) scenarios are required for comparison with observations, however the large uncertainty on $q$ found by this paper shows that we must first provide better observational constraints on all physical properties of nuclei (albedos, phase functions and shape models) to produce observed CSDs with sufficient accuracy to compare with these models.

\section*{Acknowledgments}

We thank the referee, Imre Toth, for helpful suggestions that improved this manuscript. This work was supported in part by the NASA Planetary Astronomy Program and was performed in part at the Jet Propulsion Laboratory. The research leading to these results has received funding from the European Union Seventh Framework Programme (FP7/2007-2013) under grant agreement no. 268421. This research has made use of NASA's Astrophysics Data System.

\bibliographystyle{mn2e} 
\bibliography{comets}  

\label{lastpage}

\end{document}